\documentclass[jgrga]{agutex_blank}
\pdfoutput=1
\usepackage[a4paper,textheight=698pt,textwidth=41pc]{geometry}
\usepackage{graphicx}
\setkeys{Gin}{draft=false} 

\usepackage{amsmath,amssymb,graphicx}
\usepackage{color}

\newcounter{todocount}

\usepackage{eso-pic}
\usepackage{datetime}
\newdateformat{isodate}{\THEYEAR-\twodigit{\THEMONTH}-\twodigit{\THEDAY}}



\graphicspath{{./graphics}}

\authorrunninghead{RIEMER ET AL.}  

\titlerunninghead{PARTICLE-RESOLVED MIXING STATE MODELING}

\authoraddr{N.~Riemer, Department of Atmospheric Science, University
  of Illinois at Urbana-Champaign, 105 S. Gregory St., Urbana, IL
  61801, USA. (nriemer@illinois.edu)}

\authoraddr{M.~West, Department of Mechanical Science and Engineering,
  University of Illinois at Urbana-Champaign, 1206 W. Green
  St., Urbana, IL 61801, USA. (mwest@illinois.edu)}

\authoraddr{R.~A.~Zaveri, Atmospheric Science and Global Change
  Division, Pacific Northwest National Laboratory, MSIN K9-30,
  P.O. Box 999, Richland, WA 99352, USA. (rahul.zaveri@pnl.gov)}

\authoraddr{R.~C.~Easter, Atmospheric Science and Global Change
  Division, Pacific Northwest National Laboratory, MSIN K9-30,
  P.O. Box 999, Richland, WA 99352, USA. (richard.easter@pnl.gov)}

\begin{document}

\title{Simulating the evolution of soot mixing state with a
  particle-resolved aerosol model}

\authors{N.~Riemer, \altaffilmark{1}
M.~West, \altaffilmark{2}
R.~A.~Zaveri, \altaffilmark{3}
and R.~C.~Easter\altaffilmark{3}}

\altaffiltext{1}{Department of Atmospheric Science,
University of Illinois at Urbana-Champaign,
Urbana, Illinois, USA}

\altaffiltext{2}{Department of Mechanical Science and Engineering,
University of Illinois at Urbana-Champaign,
Urbana, Illinois, USA}

\altaffiltext{3}{Atmospheric Science and Global Change
  Division, Pacific Northwest National Laboratory, Richland, Washington, USA}

\begin{abstract}
The mixing state of soot particles in the atmosphere is of crucial
importance for assessing their climatic impact, since it governs their
chemical reactivity, cloud condensation nuclei activity and radiative
properties. To improve the mixing state representation in models, we
present a new approach, the stochastic particle-resolved model
PartMC-MOSAIC, which explicitly resolves the composition of individual
particles in a given population of different types of aerosol
particles. This approach accurately tracks the evolution of the mixing
state of particles due to emission, dilution, condensation and
coagulation. To make this direct stochastic particle-based method
practical, we implemented a new multiscale stochastic coagulation
method. With this method we achieved optimal efficiency for
applications when the coagulation kernel is highly non-uniform, as is
the case for many realistic applications. PartMC-MOSAIC was applied to
an idealized urban plume case representative of a large urban area to
simulate the evolution of carbonaceous aerosols of different types due
to coagulation and condensation. For this urban plume scenario we
quantified the individual processes that contribute to the aging of
the aerosol distribution, illustrating the capabilities of our
modeling approach. The results showed for the first time the
multidimensional structure of particle composition, which is usually
lost in internally-mixed sectional or modal aerosol models.
\end{abstract}

\begin{article}

\section{Introduction}

Soot particles are an important constituent of the atmospheric
aerosol, since they participate in tropospheric chemistry
\citep{Saathoff2001} and affect human pulmonary health
\citep{Pope1996}. Because of its ability to absorb light
\citep{Horvath1993}, soot is also recognized as an important player in
the aerosol radiative forcing of climate at global, regional, and
local scales \citep{Menon2002, Chung2005, Roeckner2006}. The source of
soot particles is the incomplete combustion of carbon containing
material, which means that except for natural biomass burning all
sources of soot are of anthropogenic origin \citep{Penner1995}. The
dominant removal process is wet deposition \citep{Ducret1992}. Soot
particles can be transported over long distances reaching remote
regions such as the Arctic \citep{Clarke1985, Hansen2004}.

The initial composition of soot particles consists of black carbon and
organic carbon. The precise mixture depends heavily on the source
\citep{Medalia1982, Andreae2006}. While freshly emitted soot particles
are rather hydrophobic and present in an external mixture, their
hygroscopic qualities can change due to coagulation with soluble
aerosols, condensation of secondary organic and inorganic species, and
photochemical processes \citep{Weingartner1997}. These processes are
usually referred to as ``aging,'' and they determine the particle
growth in response to ambient relative humidity and the ability to be
activated as cloud condensation nuclei. The aging processes also have
a profound effect on the aerosol optical properties. For example,
internally mixed soot shows greater absorptivity compared to
externally mixed soot. This effect on radiative properties has been
studied by a number of investigators, e.g. \citet{Chylek1995,
  Jacobson2001, Riemer2003, Schnaiter2005, Bond2006}. Measurements
show that atmospheric soot particles are internally mixed with other
aerosol species in varying proportions, and that the hydrophobic
portion of the aerosols decreases significantly as the distance from
the sources increases \citep{Andreae1986, Levin1996, Okada2001,
  Johnson2005, Cubison2008}.

Since it is well recognized that soot particles contribute to both the
direct and indirect/semi-direct climate effect \citep{Lesins2002,
  Jacobson2000, Jacobson2002b, Nenes2002}, an adequate representation
of soot and its mixing state is sought for use in both global and
regional models, and the parameterization of soot aging is key to
determining its atmospheric abundance. Many global models have
simulated both (fresh) hydrophobic soot and (aged) hydrophilic soot,
which can be considered as a minimal representation of the soot mixing
state. Several of the models have assumed that the conversion from
hydrophobic to hydrophilic soot can be treated as an exponential decay
process, with a half-life of approximately 24~h \citep{Cooke1999,
  Lohmann1999, Koch2001, Chung2002}. This approach is a substantial
simplification since the conversion rate depends on many different
environmental conditions. This has led to more mechanistic approaches,
where processes such as condensation of sulfate on soot particles,
chemical oxidation and/or coagulation between different particle
classes are explicitly modeled to some extent \citep{Wilson2001,
  Stier2005, Tsigaridis2003}. \citet{Koch2001} and \citet{Croft2005}
compared different aging parameterizations in global models and
concluded that the model results critically depend on the respective
formulation.

To better understand the soot aging process it is desirable to have
models that are capable of representing the aerosol mixing state. From
a computational standpoint, if the aerosol mixing state can be defined
in terms of $A$ classes of chemical components (e.g., $A = 8$ with
sulfate, nitrate, ammonium, sea salt, hydrophobic organics, soluble
organics, black carbon, and mineral dust classes), then the mixing
state is an $A$-dimensional space and the size-resolved particle
composition distribution is a multivariate function.

Most existing aerosol models, however, represent the particle
population only as a bulk, or as a univariate function of a single
variable, typically total mass, diameter, or similar. To do this it is
generally assumed that the population is either fully externally
mixed, or is internally mixed with all particles in the same size bin
or mode having the same mixing state. Within this framework the
standard methods are sectional, modal, and moment models. Sectional
models (e.g. \citet{Wexler1994, Jacobson1997a, Adams1999, Zaveri2008})
place a grid on the dependent variable space and store the number
distribution or mass distribution (or both) on each grid cell. Modal
models (e.g. \citet{Whitby1991, Whitby1997, Wilson2001,Stier2005,
  Binkowski1995}) represent the particle distribution as a sum of
modes, each having a log-normal (or similar) size distribution
described by a small number of parameters (typically number, mass, and
width). Moment models (e.g. \citet{McGraw1997}) do not explicitly
resolve the distribution, but rather track a few low-order moments of
it.

It is possible to extend the standard aerosol models to handle
multivariate distributions, for example a two dimensional distribution
that is a function of two species, or a function of volume and
area. Such extensions have been investigated for sectional models
\citep{Fassi-Fihri1997}, modal models \citep{Brock1988}, and moment
models \citep{Yoon2004a, Yoon2004b}. All such models, however, require
storage and computation that scale exponentially in the number of
independent variables $A$. For the model we develop here with $A = 20$
species, fully-resolved multivariate sectional, modal, or moment
models are infeasibly expensive. For example, a sectional model
normally uses on the order of 8--20 size bins to adequately resolve a
univariate aerosol distribution, and even then will suffer from
numerical diffusion \citep{Dhaniyala1996, Wu1998}. An $A$-dimensional
distribution would thus require $8^A$--$20^A$ bins, which is
infeasible unless A is much smaller than our 20 species. In contrast,
the particle-resolved methods developed in this paper scale with the
number of particles, not the dimension of the space they are in.

While traditional univariate aerosol models are too expensive if
extended to resolve multivariate aerosol mixing states with more than
a few dimensions, there have been a number of extensions proposed to
resolve the mixing state to some extent. One example of methods that
somewhat resolve the mixing state are the so-called source oriented
models developed by \citet{Eldering1996} \citet{Kleeman1997}, and
\citet{Kleeman1998} for regional scale modeling. In these models, the
particles of different sources remain externally mixed and a number of
individual size distributions (usually about ten) are tracked, while
their mixing states change due to condensation of secondary
substances. However, because the main focus of their studies was the
prediction of particle mass size distributions, the changes in number
concentrations and particle mixing states due to self- and
hetero-coagulation of particles from different sources was
ignored. Coagulation between aerosol particles is important if one is
interested in predicting the number size distribution, especially
under polluted conditions or if long residence times are considered
\citep{Zhang2002}. Nevertheless, the source-oriented approach allows
the attribution of pollutants to specific sources and is useful for
designing emission control strategies \citep{Kleeman1998,
  Kleeman1999}. It was used in the framework of a Lagrangian
trajectory model, compared to measurements by \citet{Bhave2002}, and
has been extended to a 3D Eulerian model \citep{Kleeman2001b,
  Ying2004, Ying2006}. The Lagrangian version described in
\citet{Kleeman1998} treated the mixing state to some extent, with
fresh emissions introduced as new size distributions at every hour
along the trajectory.

Another variant of mixing state modeling was presented by
\citet{Jacobson2002}, where a total of 18 interacting aerosol size
distributions were considered. His approach was not source oriented to
the degree of Kleeman and coworkers, since anthropogenic emissions
from specific sectors were not resolved, but primary mineral dust,
sea-salt, organic matter, and black carbon were treated. Three
distributions represented black carbon at different degrees of
internal mixing.  Coagulation between different particle classes was
included, and 11 of the 18 particle classes were used to represent the
mixed particles that arise due to coagulation interaction of two
primary species.  Interactions that would result in the formation of a
particle with three different constituents resulted in a ``mixed''
particle and were not tracked further. Despite this considerable
complexity the limitation remained that particles for a certain
particle class and size were considered to be internally mixed, and
the emissions into the primary particle categories were instantly
aged.

\citet{Riemer2003} presented an approach for mixing state modeling of
soot using a mesoscale modal modeling framework. Five modes described
the composition and size distribution of sub-micron particles,
consisting of one externally mixed soot mode, two internally mixed
soot-free modes (containing inorganic and organic species), and two
internally mixed soot-containing modes. The last two modes thus
represented aged soot particles, and aging occurred either by
coagulation between modes or by condensation of secondary
substances. While this treatment allowed the distinction between fresh
and ``aged'' soot, the simplifying assumption was made that each mode
itself is internally mixed.

Here we present a particle-resolved model, PartMC, that explicitly
stores the composition of many individual aerosol particles (about
$10^5$) within a well-mixed computational volume. Relative particle
positions within this computational volume are not tracked, but rather
the coagulation process is simulated stochastically by assuming that
coagulation events are Poisson distributed with a Brownian kernel.

Applying such a Monte Carlo approach for simulating the evolution of
particle distributions dates back to \citet{Gi1975}, who developed the
exact Stochastic Simulation Algorithm (see also \citet{Gi1976},
\citet{Gi1977} and \citet{Gi1992}) to treat the stochastic
collision-coalescence process in clouds. Variants of Gillespie's
algorithm are widely used in different fields, including simulations
of gene regulatory networks \citep{SaKhPeGi2002}, chemical kinetics
\citep{Gi2007}, and sintering in flames \citep{WeMoKrWa2004}.

Since \citet{Gi1975} particle-resolved methods have been used to study
aerosols by many authors. We do not attempt to give a comprehensive
literature survey here. \citet{Ba1999} and \citet{EiWa2001b} developed
the Mass Flow Algorithm with variable computational/physical particle
ratios, \citet{KoSa2003} gave relevant error estimates, and
\citet{DeSpJo2003} coupled it to evaporation and
condensation. Somewhat similarly, \citet{LaBaDi2002} and
\citet{Alfonso2008} (based on ideas from \citet{Sp1985b}) stored the
number of particles with identical composition to reduce memory usage
and computational expense while using Gillespie's
method. \citet{Gu1997} studied convergence of stochastic coagulation
to the Smolukowski equation. \citet{EfZa2002} investigated enclosures
within aerosols using a particle-based method, while
\citet{MaKrFi2004} used particle methods with simultaneous nucleation,
coagulation, and surface growth.

While not focused on aerosol simulations, much recent work has
investigated efficient simulation methods for reaction-type Markov
processes. \citet{Gi2001} developed the tau-leaping method for
efficient generation of many events with near-constant rates, with
extensions by \citet{GiPe2003, RaPeCaGi2003, CaGiPe2005a} and others,
including for multiscale systems with scale separation in
\citet{CaGiPe2005b}. Multiscale variants of Gillespie's Stochastic
Simulation Algorithm have also been developed by
\citet{ELiVa2005b}. \citet{GiBr2000} developed the Next Reaction
Method for efficient exact sampling, which stores and reuses event
calculations for efficiency. \citet{Anderson2007} and
\citet{Anderson2008} developed efficient simulation algorithms based
on the Next Reaction Method and the tau-leaping method.

For the large number of particles in the simulations in this
paper, we used an efficient approximate coagulation method, as
described in Section~\ref{sec:coagulation} below. This used a binned
sampling method to efficiently sample from the highly multiscale
coagulation kernel in the presence of a very non-uniform particle size
distribution, implemented with a multi-event-per-timestep sampling of
the coagulation events. Multi-rate versions of Gillespie's method have
been developed previously by \citet{CaGiPe2005b} and
\citet{ELiVa2005b}, but relied on scale separation to average slow
event rates over fast timescales. The method used here does not
accelerate rare events but it does accelerate the generation of events
without scale separation, as needed for the smoothly varying
coagulation kernels and particle size distributions. The PartMC
coagulation method has storage cost proportional to the number of
physical particles, computational cost for evaporation/condensation
proportional to the number of particles, and computational cost for
coagulation proportional to the number of coagulation events.

PartMC was coupled with the new state-of-the-art aerosol chemistry
model MOSAIC \citep{Zaveri2008}, which simulates the gas- and
particle-phase chemistries, particle-phase thermodynamics, and dynamic
gas-particle mass transfer in a deterministic manner. MOSAIC treats
all the important aerosol species, including sulfate, nitrate,
chloride, carbonate, ammonium, sodium, calcium, primary organic mass
(POM), secondary organic mass, black carbon (BC), and inert organic
mass. The coupled model system, PartMC-MOSAIC, accurately predicts
number, mass, and composition size distributions, and is therefore
suited for applications where any or all of these quantities are
required.

Simulating all particles explicitly in a population of aerosol
completely eliminates any errors associated with numerical
diffusion. As a result, the highly accurate treatment of aerosol
dynamics and chemistry makes PartMC-MOSAIC suitable for use as a
numerical benchmark of mixing state for more approximate models. It
can also be applied to different environments going beyond the example
of clear-sky photochemistry shown in this paper, including the
in-cloud processing of aerosol, and it can be used to accurately
estimate quantities that depend on the mixing state, such as cloud
condensation nuclei spectra and optical properties, which we will
address in a forthcoming paper. The current version of PartMC is
available under the GNU General Public License (GPL) at
http://www.mechse.uiuc.edu/research/mwest/partmc/, while the MOSAIC
code is available upon request from R.~A.~Zaveri.

This manuscript is organized as follows. In
Section~\ref{sec:true_eqns} we write the governing equations for the
coupled gas-aerosol box model and discuss the approximations needed by
this model of the physical system. The numerical approximation to the
governing equations is given in Section~\ref{sec:particle_models},
where we introduce the particle-resolved aerosol model PartMC and
describe how it is coupled to the gas- and aerosol-chemistry code
MOSAIC. In Section~\ref{sec:coagulation} we give the efficient
coagulation algorithm used by PartMC and verify its performance
numerically. Finally, Section~\ref{sec:urban_plume} demonstrates the
capabilities of this new model approach by focusing on the evolution
of the mixing state of soot particles in an idealized urban plume
scenario. The main contributions of this paper are: 1) an accelerated
stochastic coagulation method for multiscale kernels, 2) the coupling
of a particle resolved model with a gas- and aerosol-chemistry code,
and 3) an initial study of the soot mixing states present in a typical
polluted urban environment.

\section{Coupled aerosol-gas governing equations}
\label{sec:true_eqns}

We consider a Lagrangian parcel framework where we simulate the
evolution of aerosol particles and trace gases in single parcel (or
volume) of air moving along a specified trajectory. In addition to
coagulation and aerosol and gas chemistry, the model treats prescribed
emissions of aerosols and gases, and mixing of the parcel with
``background air''.

The evolution of atmospheric aerosols is extremely complex, involving
interactions between fluid-transport and micro-scale properties of the
aerosol particles during coagulation and condensation. A fully
resolved simulation of fluid-aerosol interaction cannot capture a
large enough system to determine macro-scale aerosol distribution
properties. Instead of fully resolved models, it is usual to use a box
model locally and store the size distribution of aerosol in a certain
physical volume without storing the positions of the particles in
three dimensions. Particle interactions such as coagulation are then
represented stochastically by a kernel $K$ that defines coagulation
probability rates for particles depending on their sizes and
compositions. This representation assumes that the mixing timescale is
significantly faster than the coagulation and condensation
timescales. Equivalently, we are assuming that the aerosol processes
are almost Markovian and so a memoryless model is a good approximation
to the physics. This assumption underlies all sectional and modal
models in use today, as well as our particle-resolved model PartMC,
and means that we need only consider the aerosol- and gas-species
densities.

An aerosol particle contains mass $\mu_a \ge 0$ of species $a$, for $a
= 1,\ldots,A$, so that the particle composition is described by the
$A$-dimensional vector $\vec{\mu} \in \mathbb{R}^A$.  $\mu_{\rm all}$
($\rm m^3$) is the total wet mass of the particle, and $\mu_{\rm dry}
= \mu_{\rm all} - \mu_{\rm H_2O}$ ($\rm m^3$) is the total dry mass.
The cumulative aerosol number distribution at time $t$ and constituent
masses $\vec{\mu} \in \mathbb{R}^A$ is $N(\vec{\mu},t)$ ($\rm
m^{-3}$), which is defined to be the number density of aerosol
particles that contain less than $\mu_a$ mass of species $a$, for all
$a = 1,\ldots,A$. The aerosol number distribution at time $t$ and
constituent masses $\vec{\mu} \in \mathbb{R}^A$ is $n(\vec{\mu},t)$
($\rm m^{-3}\,kg^{-A}$), which is defined by
\begin{equation}
n(\vec{\mu},t) = \frac{\partial^A N(\vec{\mu}, t)}{\partial \mu_1
\partial \mu_2 \ldots \partial \mu_A}.
\end{equation}

The concentration of trace gas phase species $i$ at time $t$ is given
by $g_i(t)$ ($\rm mol\,m^{-3}$), for $i = 1,\ldots,G$, so the trace
gas phase species concentrations are the $G$-dimensional vector
$\vec{g}(t) \in \mathbb{R}^G$. We assume that the aerosol and gas
species are numbered so that the first $C$ species of each undergo gas
to particle conversion, and that they are in the same order so that
gas species $i$ converts to aerosol species $i$, for $i = 1,\ldots,C$.

The environment is described by temperature $T(t)$ ($\rm K$), pressure
$p(t)$ ($\rm Pa$), relative humidity ${\rm RH}(t)$ (dimensionless),
and dry density $\rho_{\rm dry}(t)$ ($\rm kg\,m^{-3}$).  For the
simulation in Section~\ref{sec:urban_plume} the air temperature is
prescribed as a function of time, while the air pressure and water
mixing ratio are kept constant and the relative humidity and dry
density are updated accordingly.

We assume that we are modeling a vertical slice of a well-mixed
boundary layer during the day and a slice of the residual layer during
the night, always surrounded to the sides and above by background
air. The height of the boundary layer is given by $H(t)$ ($\rm m$). We
denote by $\lambda_{\rm dil,horiz}(t)$ ($\rm s^{-1}$) the horizontal
dilution rate with the prescribed background gas and aerosol, and by
$\lambda_{\rm dil,vert}(t)$ ($\rm s^{-1}$) the vertical dilution rate
that represents entrainment of a growing boundary layer. The total
dilution rate $\lambda_{\rm dil}^{\rm eff}(t)$ ($\rm s^{-1}$) is then
given by
\begin{align}
\label{eqn:dil}
\lambda_{\rm dil}(t)
&= \lambda_{\rm dil,horiz}(t) + \lambda_{\rm dil,vert}(t) \\
\lambda_{\rm dil,vert} &=  I_{\rm entrain}(t) \max\biggl(0,
\frac{1}{H(t)} \frac{dH(t)}{dt}\biggr),
\end{align}
where vertical entrainment only occurs for increasing boundary layer
heights. The indicator $I_{\rm entrain}(t)$ is $1$ when the modeled
air parcel is within the boundary layer and so entrainment is
possible, and is $0$ when the air parcel is in the residual layer.

To obtain the mean evolution in the large-number limit we neglect
correlations between the number of particles of different sizes
\citep{Gi1972} and thus obtain the classical Smoluchowski coagulation
equation \citep{Sm1916a, Sm1916b}, which for a multidimensional
aerosol distribution with gas coupling is
\begin{subequations}
\label{eqn:main}
\begin{align}
  \frac{\partial n(\vec{\mu},t)}{\partial t} &= \frac{1}{2} \int_0^{\mu_1}
  \int_0^{\mu_2} \cdots \int_0^{\mu_A} K(\vec{\mu}', \vec{\mu} - \vec{\mu}')
  \nonumber \\
  & \hspace{4em} \times n(\vec{\mu}',t) \, n(\vec{\mu} -
  \vec{\mu}',t)\, d\mu_1' \, d\mu_2' \ldots d\mu_A'
  \label{eqn:main_aero_coag_gain} \\
  &\qquad - \int_0^\infty \int_0^\infty \cdots \int_0^\infty K(\vec{\mu},
  \vec{\mu}') \nonumber \\
  & \hspace{4em} \times n(\vec{\mu},t) \, n(\vec{\mu}',t)\,
  d\mu_1' \, d\mu_2' \ldots d\mu_A'   \label{eqn:main_aero_coag_loss}
  \displaybreak[0] \\
  &\qquad + \dot{n}_{\rm emit}(\vec{\mu},t) \label{eqn:main_aero_emit} \\
  &\qquad + \lambda_{\rm dil}(t)
  \Bigl(n_{\rm back}(\vec{\mu},t) - n(\vec{\mu},t)\Bigr)
  \label{eqn:main_aero_dilute} \displaybreak[0] \\
  &\qquad - \sum_{i = 1}^C \frac{\partial}{\partial
    \mu_i} \biggl(c_i I_i(\vec{\mu}, \vec{g}, t)
  \,n(\vec{\mu},t) \biggr) \label{eqn:main_aero_gas_transfer} \\
  &\qquad - \frac{\partial}{\partial
    \mu_{C+1}} \biggl(c_{\rm w} I_{\rm w}(\vec{\mu}, \vec{g}, t)
  \,n(\vec{\mu},t) \biggr) \label{eqn:main_aero_water_transfer} \\
  &\qquad + \frac{1}{\rho_{\rm dry}(t)} \frac{d \rho_{\rm dry}(t)}{dt}
  n(\vec{\mu},t) \label{eqn:main_aero_temp_press} \displaybreak[0] \\
  \frac{dg_i(t)}{dt} &= \dot{g}_{{\rm emit},i}(t)
  + \lambda_{\rm dil}(t) \Bigl(g_{{\rm back},i}(t) - g_i(t)\Bigr)
  \label{eqn:main_gas_emit_dilute} \\
  &\qquad + R_i(\vec{g}) \label{eqn:main_gas_react} \\
  &\qquad - \int_0^\infty \int_0^\infty \cdots \int_0^\infty
  I_i(\vec{\mu}, \vec{g}, t) \nonumber \\
  & \hspace{8em} \times n(\vec{\mu},t) \, d\mu_1 \, d\mu_2 \ldots d\mu_A
  \label{eqn:main_gas_aero_transfer} \\
  &\qquad 
  + \frac{1}{\rho_{\rm dry}(t)} \frac{d \rho_{\rm dry}(t)}{dt}
  g_i(t) \label{eqn:main_gas_temp_press}.
\end{align}
\end{subequations}
The integro-differential equation~(\ref{eqn:main}) must be augmented
with appropriate boundary conditions. This is done on physical grounds
to ensure that the constituent masses of particles cannot become
negative. In equation~(\ref{eqn:main}), $K(\vec{\mu}_1,\vec{\mu}_2)$
($\rm m^3\,s^{-1}$) is the coagulation rate between particles with
constituent masses $\vec{\mu}_1$ and $\vec{\mu}_2$, $\dot{n}_{\rm
  emit}(\vec{\mu},t)$ ($\rm m^{-3}\,kg^{-A}\,s^{-1}$) is the number
distribution rate of aerosol emissions, $n_{\rm back}(\vec{\mu},t)$
($\rm m^{-3}\,kg^{-A}$) is the background number distribution,
$I_i(\vec{\mu}, \vec{g}, t)$ ($\rm mol\,s^{-1}$) is the condensation
or evaporation flux of gas species $i$ (with $I_{\rm w}(\vec{\mu},
\vec{g}, t)$ the flux for water), $c_i$ ($\rm m^3 \, mol^{-1}$) is the
conversion factor from moles of gas species $i$ to mass of aerosol
species $i$ (with $c_{\rm w}$ the factor for water), $\dot{g}_{{\rm
    emit},i}(t)$ ($\rm mol\,m^{-3}\,s^{-1}$) is the emission rate of
gas species $i$, $g_{{\rm back},i}(t)$ ($\rm mol\,m^{-3}$) is the
background concentration of gas species $i$, and $R_i(\vec{g})$ ($\rm
mol\,m^{-3}\,s^{-1}$) is the concentration growth rate of gas species
$i$ due to gas chemical reactions. Many of the rates, coefficients and
functions also depend on the environmental conditions, but we have not
written this dependence explicitly.

\section{Particle-resolved aerosol models}
\label{sec:particle_models}

\subsection{PartMC aerosol state representation}

We consider a Lagrangian parcel with volume $V$ ($\rm m^3$), also
called the computational volume. We represent the aerosol state by
storing $N_{\rm MC}$ particles in this volume, written $\Pi =
(\vec{\mu}^1,\vec{\mu}^2,\ldots,\vec{\mu}^{N_{\rm MC}})$, where the
particle order is not significant. Each particle is an $A$-dimensional
vector $\vec{\mu}^i \in \mathbb{R}^A$ with components
$(\mu^i_1,\mu^i_2,\ldots,\mu^i_A)$, so $\mu^i_a$ is the mass of
species $a$ in particle $i$, for $a = 1,\ldots,A$ and $i =
1,\ldots,N_{\rm MC}$. In the notation of \citet{DeSpJo2003} for the
Mass Flow Algorithm, we are taking $(\omega_i/y_i)(t) = 1$, which
means one computational particle per physical particle. While we track
every particle within the computational volume $V$, we regard this
volume as being representative of a much larger air parcel. For
example, in Section~\ref{sec:urban_plume} we use a computational
volume on the order of a few cubic centimeters but take this to be
approximating the state of the well-mixed boundary layer during the
day and the residual layer during the night.

The simulation of the aerosol state proceeds by two mechanisms. First,
the composition of each particle can change, changing the components
of the vector $\vec{\mu}^i$ as species condense from the gas phase and
evaporate to it, for example. Second, the population $\Pi$ can have
particles added and removed, either by emissions, dilution or
coagulation events between particles.

The representation of the aerosol as a finite collection of particles
$\Pi$ in a volume $V$ is very flexible, as other properties can easily
be stored for each particle, such as fractal dimension, electric
charge, age since emission, etc. In the present paper we store the
number of coagulation events undergone by each particle to produce
Figure~\ref{fig:aero_2d_n_orig}.

\subsection{PartMC emissions}
\label{sec:emission_dilution}

Because we are using a finite number of particles to approximate the
current aerosol population, we need to add a finite number of emitted
particles to the volume at each timestep. Over time these finite
particle samplings should approximate the continuum emission
distribution, so the samplings at each timestep must be different. We
assume that emissions are memoryless, so that emission of each
particle is uncorrelated with emission of any other particle. Under
this assumption the appropriate statistics are Poisson distributed,
whereby the distribution of finite particles is parametrized by the
mean emission rate and distribution.

Consider a number distribution production rate $\dot{n}_{\rm
  emit}(\vec{\mu}, t)$ ($\rm m^{-3}\,kg^{-A}\,s^{-1}$), a volume $V$
($\rm m^3$), and a timestep $\Delta t$ ($\rm s$). The emissions over
the timestep from time $t_0$ to $t_1 = t_0 + \Delta t$ are given by
\begin{align}
\label{eqn:integrated_emissions}
n_{\rm emit}(\vec{\mu}; t_0, t_1)
&= \int_{t_0}^{t_1} \dot{n}_{\rm emit}(\vec{\mu}, t) \, dt \\
&\approx (t_1 - t_0) \dot{n}_{\rm emit}(\vec{\mu}, t_0)
\end{align}
for which we use the first-order approximation above.  To obtain a
finite Poisson sample of the distribution $n(\vec{\mu}) = n_{\rm
  emit}(\vec{\mu}; t_0, t_1)$ ($\rm m^{-3}\,kg^{-A}$) in the
computational volume $V$ we first see that the mean number $N(n, V)$
of sampled particles will be
\begin{equation}
\label{eqn:poisson_particles_mean}
N_{\rm mean}(n, V) = \int_0^\infty \int_0^\infty \cdots \int_0^\infty
n(\vec{\mu}) V \, d\mu_1\,d\mu_2\ldots d\mu_A.
\end{equation}
The actual number $S$ of emitted particles added in a timestep will be
Poisson distributed, written $S \sim \operatorname*{Pois}(\lambda)$,
for mean $\lambda = N_{\rm mean}(n, V)$, so that
\begin{equation}
\label{eqn:poisson}
\operatorname*{Prob}(S = k) = \frac{\lambda^k e^{-\lambda}}{k!} \text{
  for } k \in \mathbb{Z}^+.
\end{equation}
A Poisson sampling $\Pi_{\rm samp}$ of the number distribution
$n(\vec{\mu})$ in volume $V$, written $\Pi_{\rm samp} \sim
\operatorname*{Pois_{dist}}(n, V)$, is a finite sequence of particles
given by
\begin{subequations}
\label{eqn:pois_dist}
\begin{align}
\label{eqn:pois_dist_particles}
\Pi_{\rm samp} &= (\vec{\mu}^1,\vec{\mu}^2,\ldots,\vec{\mu}^S) \\
\label{eqn:pois_dist_number}
S &\sim \operatorname*{Pois}\Bigl(N_{\rm mean}(n, V)\Bigr) \\
\label{eqn:pois_dist_sample_particle}
\vec{\mu}^s &\sim \frac{n(\vec{\mu}) V}{N_{\rm mean}(n, V)}
\text{ for } s = 1,\ldots,S,
\end{align}
\end{subequations}
where~(\ref{eqn:pois_dist_sample_particle}) means that each particle
has a composition drawn from the distribution specified by
$n(\vec{\mu})$.

\subsection{PartMC dilution}

As with emissions, we must also obtain a finite sampling of background
particles that have diluted into our computational volume during each
timestep. In addition, some of the particles in our current sample
will dilute out of our volume and will be lost, so this must be
sampled as well. We assume that dilution is memoryless, so that
dilution of each particle is uncorrelated with the dilution of any
other particle or itself at other times, and that once a particle
dilutes out it is no more likely to re-enter than any other background
particle.

Let the background particle distribution be $n_{\rm back}(\vec{\mu},
t)$ ($\rm m^{-3}\,kg^{-A}$), the computational volume be $V$ ($\rm
m^3$), and the timestep be $\Delta t$ ($\rm s$). The distribution of
particles that dilute from the background into the volume $V$ between
times $t_0$ and $t_1 = t_0 + \Delta t$ is $n_{\rm dil}(\vec{\mu}; t_0,
t_1)$, where $n_{\rm dil}(\vec{\mu}; t_0, t)$ satisfies
\begin{subequations}
\begin{align}
  \frac{\partial n_{\rm dil}(\vec{\mu}; t_0, t)}{\partial t}
  &= \lambda_{\rm dil}(t)
  \Bigl(n_{\rm back}(\vec{\mu},t) - n_{\rm dil}(\vec{\mu}; t_0, t)\Bigr) \\
  n_{\rm dil}(\vec{\mu}; t_0, t_0) &= 0
\end{align}
\end{subequations}

We use the first-order approximation given by
\begin{align}
  n_{\rm dil}(\vec{\mu}; t_0, t_1)
  &\approx (t_1 - t_0) \, \lambda_{\rm dil}(t_0) \, n_{\rm back}(\vec{\mu},t_0)
\end{align}
A discrete sampling of $n_{\rm dil}(\vec{\mu}; t_0, t_1)$ is then
given by $\Pi_{\rm dil} \sim {\rm Pois}_{\rm dist}(n_{\rm
  dil}(\vec{\mu}), V)$, as in~(\ref{eqn:pois_dist}).

If we start the timestep at time $t_0$ with the particle population
$\Pi$, then each particle in $\Pi$ has probability $p(t_0, t_1)$ to be
lost by dilution during the timestep, where $p(t_0, t)$ satisfies
\begin{subequations}
\begin{align}
\frac{\partial p(t_0, t)}{\partial t}
&= \lambda_{\rm dil}(t) \Bigl(1 - p(t_0, t)\Bigr) \\
p(t_0, t_0) &= 0
\end{align}
\end{subequations}
We use the first-order approximation given by
\begin{equation}
p(t_0, t_1) \approx (t_1 - t_0) \lambda_{\rm dil}(t_0)
\end{equation}
We denote the binomial distribution for number $n$ and probability $p$
by $B(n,p)$. The number of particles lost from $\Pi$ between times
$t_0$ and $t_1 = t_0 + \Delta t$ is then given by $N_{\rm loss}$ which
is distributed as
\begin{align}
N_{\rm loss} &\sim B\Bigl(N_{\rm MC}, p(t_0, t_1)\Bigr) \\
&\approx {\rm Pois}(N_{\rm MC} \, p(t_0, t_1)\Bigr)
\end{align}
We approximate the binomial distribution with a Poisson distribution
as above, which converges as $\Delta t \to 0$ for fixed $N_{\rm
  MC}$. As each particle has equal probability to be lost due to
dilution, we can sample $N_{\rm loss}$ and then choose $N_{\rm loss}$
particles uniformly from $\Pi$ to be removed.

\subsection{Coupled PartMC-MOSAIC method}
\label{sec:coupled_partmc_mosaic}

\begin{figure}
  \flushleft \newlength{\innerboxwidthB}
  \setlength{\innerboxwidthB}{\columnwidth}
  \addtolength{\innerboxwidthB}{-2\fboxsep}
  \noindent
  \framebox{\begin{minipage}{\innerboxwidthB}
      \flushleft
      \begin{list}{}{\setlength{\leftmargin}{2em}\setlength{\itemindent}{-2em}\setlength{\labelsep}{0pt}}
      \item $\Pi$ is the sequence of particle compositions
      \item $V$ is the computational volume
      \item $\vec{g}$ is the gas concentrations
      \item $t = 0$
      \item \textbf{while} $t < t_{\rm final}$ \textbf{do:}
        \begin{list}{}{\setlength{\leftmargin}{3em}\setlength{\itemindent}{-2em}}
        \item $t = t + \Delta t$
        \item update temperature $T(t)$, pressure $p(t)$, relative
          humidity $RH(t)$, dry density $\rho_{\rm dry}(t)$, and
          mixing height $H(t)$
        \item $V(t) = V(t - \Delta t)
          \frac{\rho_{\rm dry}(t - \Delta t)}{\rho_{\rm dry}(t)}$
        \item $\vec{g}(t) = \vec{g}(t - \Delta t)
          \frac{\rho_{\rm dry}(t)}{\rho_{\rm dry}(t - \Delta t)}$
        \item add $\Delta t \, \dot{\vec{g}}_{\rm emit}(t) + \Delta t \,
          \lambda_{\rm dil}(t) \Bigl(\vec{g}_{\rm back}(t)
          - \vec{g}(t)\Bigr)$ to $\vec{g}$
        \item randomly choose $N_{\rm loss} \sim
          \operatorname*{Pois}(\Delta t \, \lambda_{\rm dil} N_{\rm MC})$ and
          remove $N_{\rm loss}$ randomly chosen particles from
          $\Pi$
        \item add a sample of $\operatorname*{Pois_{dist}}\Bigl(
          \lambda_{\rm dil} \, \Delta t \, n_{\rm back}(\cdot,
          t), V\Bigr)$ to $\Pi$
        \item add a sample of $\operatorname*{Pois_{dist}}\Bigl(
          \Delta t \, \dot{n}_{\rm emit}(\cdot, t), V\Bigr)$ to $\Pi$
        \item perform one $\Delta t$-timestep of coagulation for
          $\Pi$ with the PartMC algorithm in
          Figure~\ref{fig:partmc_coag_alg}
        \item integrate the system of coupled
          ODEs~(\ref{eqn:main_aero_gas_transfer})
          and~(\ref{eqn:main_gas_react})--(\ref{eqn:main_gas_aero_transfer})
          with MOSAIC for time $\Delta t$
        \end{list}
      \item \textbf{end while}
      \end{list}
  \end{minipage}}
  \caption{Coupled PartMC-MOSAIC algorithm. \label{fig:partmc_mosaic_alg}}
\end{figure}

We couple the stochastic PartMC particle-resolved aerosol model to the
deterministic MOSAIC gas- and aerosol-chemistry code in a
time-splitting fashion to obtain a complete discretization of the
governing equations~(\ref{eqn:main}). The aerosol distribution
$n(\vec{\mu},t)$ is represented by $N_{\rm MC}$ particles in a
computational volume $V$, as described above, while the gas vector
$\vec{g}(t)$ stores the gas concentrations.
Equations~(\ref{eqn:main_aero_coag_gain})--(\ref{eqn:main_aero_dilute})
are solved stochastically by the PartMC code, and the gas
equations~(\ref{eqn:main_gas_emit_dilute})--(\ref{eqn:main_gas_aero_transfer})
together with the coupling term~(\ref{eqn:main_aero_gas_transfer}) are
integrated deterministically by the MOSAIC code. The
terms~(\ref{eqn:main_aero_temp_press})
and~(\ref{eqn:main_gas_temp_press}) are implemented deterministically
by updating $V$ and $\vec{g}$ by the density change or its inverse, as
appropriate. The full coupled PartMC-MOSAIC algorithm is given in
Figure~\ref{fig:partmc_mosaic_alg}.

The current version of MOSAIC treats all the locally and globally
important aerosol species including SO$_4$, NO$_3$, Cl, CO$_3$, MSA
(methanesulfonic acid), NH$_4$, Na, Ca, other inorganic mass, BC, and
POM, and secondary organic mass. It consists of four computationally
efficient modules: 1) the gas-phase photochemical mechanism CBM-Z
\citep{Zaveri1999}; 2) the Multicomponent Taylor Expansion Method
(MTEM) for estimating activity coefficients of electrolytes and ions
in aqueous solutions \citep{Zaveri2005a}; 3) the Multicomponent
Equilibrium Solver for Aerosols (MESA) for intra-particle solid-liquid
partitioning \citep{Zaveri2005b}; and 4) the Adaptive Step Time-split
Euler Method (ASTEM) for dynamic gas-particle partitioning over size-
and composition-resolved aerosol \citep{Zaveri2008}. The version of
MOSAIC box-model implemented here also includes a treatment for
secondary organic aerosol (SOA) based on the SORGAM scheme
\citep{Schell2001}.

\vspace{3em} 

\section{PartMC coagulation algorithm}
\label{sec:coagulation}

\subsection{Stochastic coagulation simulation}
\label{sec:gillespie}

If we have $N_{\rm MC}$ particles then there are $N_{\rm MC}(N_{\rm
  MC}-1)/2$ possible coagulation events, with the probability rate of
a coagulation between particles $i$ and $j$ in a volume $V$ given by
$K(\vec{\mu}^i,\vec{\mu}^j) / V$ for the coagulation kernel
$K(\vec{\mu}^i,\vec{\mu}^j)$ ($\rm m^{3}\,s^{-1}$). The only
difficulty with stochastic coagulation is generating a sequence of
coagulation events, each consisting of a pair of particles $(i,j)$
that coagulate and a time $\Delta t$ until the coagulation
occurs. Once a coagulation event is determined then we simply remove
the particles $i$ and $j$ from the population $\Pi$, add a new
particle with composition $\vec{\mu}^{\rm new} = \vec{\mu}^i +
\vec{\mu}^j$, and advance the time by $\Delta t$.

The standard stochastic simulation algorithm for this system is due to
\citet{Gi1975} and is based on the observation that the probability
density for the time until the next coagulation event is
\begin{equation}
\label{eqn:p_next_t}
P(\Delta t) = \frac{K_{\text{tot}}}{V} e^{-K_{\text{tot}} \Delta t/V},
\end{equation}
where $K_{\text{tot}} = \sum_{i<j} K(\vec{\mu}^i,\vec{\mu}^j)$ is the
total rate. We can thus generate an elapsed time by sampling the
probability density function~(\ref{eqn:p_next_t}). The conditional
probability that the coagulation event that occurred was between
particles $i$ and $j$ is then
\begin{equation}
\label{eqn:p_ij_given_t}
P(i,j \mid \Delta t) = \frac{K(\vec{\mu}^i,\vec{\mu}^j)}{K_{\text{tot}}},
\end{equation}
and this can be sampled to determine which particles coagulated, and
then the coagulation event can be performed.

Gillespie's method has the advantage that it generates exact
realizations of the stochastic coagulation process. It faces two main
difficulties in practice, however. First, the total rate
$K_{\text{tot}}$ continually changes as coagulation events occur and
particle compositions change due to condensation. Computing a
reasonable estimate of this parameter quickly becomes exceedingly
expensive, and approximations made to speed up this estimate introduce
errors that are difficult to estimate and control. Second, while
sampling~(\ref{eqn:p_next_t}) is very cheap,
sampling~(\ref{eqn:p_ij_given_t}) can be expensive for complex
kernels. The two main methods are use of the cumulative distribution
function, which scales badly in the number of particles and is thus
too expensive for large particle numbers, and use of
accept-reject. While accept-reject scales well as the number of
particles grows, it is very inefficient if the kernel $K_{ij}$ is
highly non-uniform, as is unfortunately the case for many physically
relevant aerosol distributions. Despite these difficulties,
Gillespie's method is by far the most commonly used method in
practice, with many slight variants appearing in the literature (for
example, see \citet{EfZa2002, KrMaFi2000, GaBrAeSe1987, FiWe1991}).

To avoid these two difficulties we formulate an improved method. We
use a fixed timestep method and we develop a binned acceptance
procedure. The use of a fixed timestep removes the need to know
$K_{\text{tot}}$, albeit with the introduction of some error. This
fixed timestep also makes it easy to integrate the coagulation with
other physics and chemistry using a time-splitting scheme. The binned
sampling method means that we are not subject to slow-downs from
non-uniform kernels.

\subsection{Fixed-timestep stochastic coagulation}

We choose a fixed timestep $\Delta t$ and in each timestep choose
$N_{\text{test}}$ particle pairs to test. We then generate
$N_{\text{test}}$ random particle pairs uniformly and for each pair
$(i,j)$ we accept a coagulation event with probability
\begin{align}
P(i,j) &= 1 - \exp\biggl(\frac{-K(\vec{\mu}^i,\vec{\mu}^j) \Delta t}{V}
\frac{N_{\rm MC}(N_{\rm MC}-1)}{N_{\text{test}}}\biggr) \nonumber \\
\label{eqn:fixed_P}
& \approx \frac{K(\vec{\mu}^i,\vec{\mu}^j) \Delta t}{V}
\frac{N_{\rm MC} (N_{\rm MC}-1)}{N_{\text{test}}}.
\end{align}
In the limit $\Delta t \to 0$ this generates an exact realization of
the stochastic coagulation process, and for finite $\Delta t$
introduces a discretization error. The number $N_{\text{test}}$ should
be chosen large enough that $P(i,j) \le 1$ for all pairs $i,j$ and for
convergence it must remain bounded away from zero as $\Delta t \to
0$. This is similar to the sampling technique used in
\citet{DeSpJo2003}.

The efficiency of the method, as with any procedure of accept-reject
type, is greatest when the maximum value of $P(i,j)$ is as close as
possible to $1$. To ensure this we choose
\begin{equation}
  \label{eqn:fixed_N_c}
  N_{\rm test} = \Bigl\lceil K_{\text{max}}
  \Delta t N_{\rm MC}(N_{\rm MC}-1) / V \Bigr\rceil,
\end{equation}
where $K_{\text{max}} = \max_{i,j} K(\vec{\mu}^i,\vec{\mu}^j)$ is the
maximum kernel value and $\lceil x \rceil$ is the least integer
greater than $x$. In practice we take $K_{\text{max}}$ to be a cheaply
computable upper-bound for $K(\vec{\mu}^i,\vec{\mu}^j)$, which
slightly increases the accuracy of the method and is much cheaper.

The fixed timestep method thus cleanly resolves the difficulties with
Gillespie's method to do with the need to determine
$K_{\text{tot}}$. It still has the problem, however, that if the
kernel $K(\vec{\mu}^i,\vec{\mu}^j)$ is very non-uniform then the
acceptance procedure will be very inefficient. To fix this, we adopt a
binned approach.

\subsection{Binned stochastic coagulation}
\label{sec:partmc}

For coagulation kernels of physical interest, such as those arising
from Brownian motion or gravitational settling, the kernel
$K(\vec{\mu}, \vec{\mu}')$ is highly multiscale, with many orders of
magnitude difference between the highest and lowest rates. This is a
problem for the sampling procedure outlined in the previous section,
because $N_{\rm test}$ will be very large and so we will have to
reject many events using~(\ref{eqn:fixed_P}) for each accepted event.

\begin{figure}
\flushleft
\newlength{\innerboxwidthA}
\setlength{\innerboxwidthA}{\columnwidth}
\addtolength{\innerboxwidthA}{-2\fboxsep}
\noindent
\framebox{\begin{minipage}{\innerboxwidthA}
  \begin{list}{}{\setlength{\leftmargin}{2em}\setlength{\itemindent}{-2em}\setlength{\labelsep}{0pt}}
  \item divide diameter axis into bins as for a sectional model
  \item $N_{\rm MC}(b)$ is the number of particles in bin $b$
  \item $\vec{\mu}(b,i)$ is the mass vector of the $i$-th particle in bin $b$
  \item $K_{\text{max}}(b_1,b_2)$ is a precomputed upper bound on the
    kernel for any particles from bins $b_1$ and $b_2$
  \item $\Delta t$ is the timestep
  \item \textbf{for all} bin pairs $(b_1,b_2)$ \textbf{do:}
    \begin{list}{}{\setlength{\leftmargin}{2.5em}\setlength{\itemindent}{-1em}}
    \item $N_{\rm event} = N_{\rm MC}(b_1) N_{\rm MC}(b_2) / 2$
    \item $N_{\text{test}} = \Bigl\lceil K_{\text{max}}(b_1,b_2) \, \Delta t \, N_{\rm event} / V \Bigr\rceil$
    \item \textbf{for} $N_{\text{test}}$ repetitions \textbf{do:}
      \begin{list}{}{\setlength{\leftmargin}{2.5em}\setlength{\itemindent}{-1em}}
      \item randomly choose particles $i_1$ and $i_2$ uniformly in
        bins $b_1$ and $b_2$
      \item $K_{12} = K\bigl(\vec{\mu}(b_1,i_1),\vec{\mu}(b_2,i_2)\bigr)$
      \item randomly choose $r$ uniformly in $[0,1]$
      \item \textbf{if} $r < K_{12} \, \Delta t \, N_{\rm event} /
        (N_{\text{test}} V)$ \textbf{then:}
        \begin{list}{}{\setlength{\leftmargin}{2.5em}\setlength{\itemindent}{-1em}}
        \item coagulate the two particles, updating the arrays $N(b)$
          and $\vec{\mu}(b,i)$
        \end{list}
      \item \textbf{end if}
      \end{list}
    \item \textbf{end for}
    \end{list}
  \item \textbf{end for}
  \end{list}
\end{minipage}}
  \caption{PartMC coagulation algorithm. \label{fig:partmc_coag_alg}}
\end{figure}

To accelerate this procedure we take advantage of the fact that the
kernel $K(\vec{\mu}^i,\vec{\mu}^j)$ is not random in its
non-uniformity, but rather depends primarily on the diameter of the
particles. This means that if pairs $i,j$ and $k,\ell$ are similar, so
that the diameters of particles $i$ and $k$ are close, as are the
diameters of particles $j$ and $\ell$, then
$K(\vec{\mu}^i,\vec{\mu}^j) \approx K(\vec{\mu}^k,\vec{\mu}^\ell)$. We
thus group particles into bins sorted by diameter and we use the
acceptance procedure~(\ref{eqn:fixed_P}) for each pair of bins
separately. This binned approach ensures that all particle pairs under
consideration in a particular iteration have similar coagulation
rates, and hence the procedure will have a high proportion of
acceptances. Use of a binned version of the fixed timestep algorithm
means that the number of samples~(\ref{eqn:fixed_N_c}) done per pair
of bins is automatically adapted to the number of particles in those
bins. It also allows us to pre-compute the $K_{\text{max}}$ values for
each bin pair. The resulting algorithm is shown in
Figure~\ref{fig:partmc_coag_alg}.

The primary disadvantage of using a binned sampling procedure is in
code complexity, as the bin structures of particles with similar sizes
need to be constructed and maintained. This also adds a small amount
of computational overhead to the coagulation routine, which is far
outweighed by the enormous efficiency gains. We should note that the
binned sampling procedure introduces no error in the simulation and is
a pure efficiency gain. For typical aerosol profiles the binned
procedure gives about two to four orders-of-magnitude speedup in
computational time, as quantified in Section~\ref{sec:verification}.

As a result of continuous coagulation the total number of simulated
particles decreases with time. To maintain sufficient resolution and
adequate statistics we double the number of particles whenever the
number of particles becomes less than half of the original particle
number. This corresponds to a doubling of the computational
volume. When we include emissions of particles, the particle number
eventually becomes too large and we run into computational limits. To
avoid this we halve the number of particles when we have more than
twice the original number of particles, which corresponds to a halving
of the computational volume.

For some kernels, such as the Brownian kernel used in
Section~\ref{sec:urban_plume}, the kernel is primarily dependent on
the particle diameters but also depends on particle density. We could
store the particles sorted into a 2D array per-diameter-bin and
per-density-bin, but the density variation is bounded and small enough
that it is still reasonably efficient to store them only
per-diameter-bin and to compute $K_{\rm max}$ to take the maximum
density variation into account.

To enable efficient coagulation, the particle array $\Pi$ is stored as
an array of pointers to partially-filled particle arrays, one per
diameter-bin. Insertions into bin arrays are performed at the end of
the currently filled area and deletions from the middle are followed
by a shift of the last element into the gap, ensuring full packing of
each bin array at all times. Each diameter-bin array is reallocated to
twice its existing size when necessary or half its existing size when
possible. This gives constant-time random access at the cost of
$O(\log \Delta N_{\rm MC})$ reallocations and at most twice the
minimal memory usage.

\subsection{Verification of the PartMC coagulation algorithm}
\label{sec:verification}

\begin{figure}
  \begin{center}
    \includegraphics{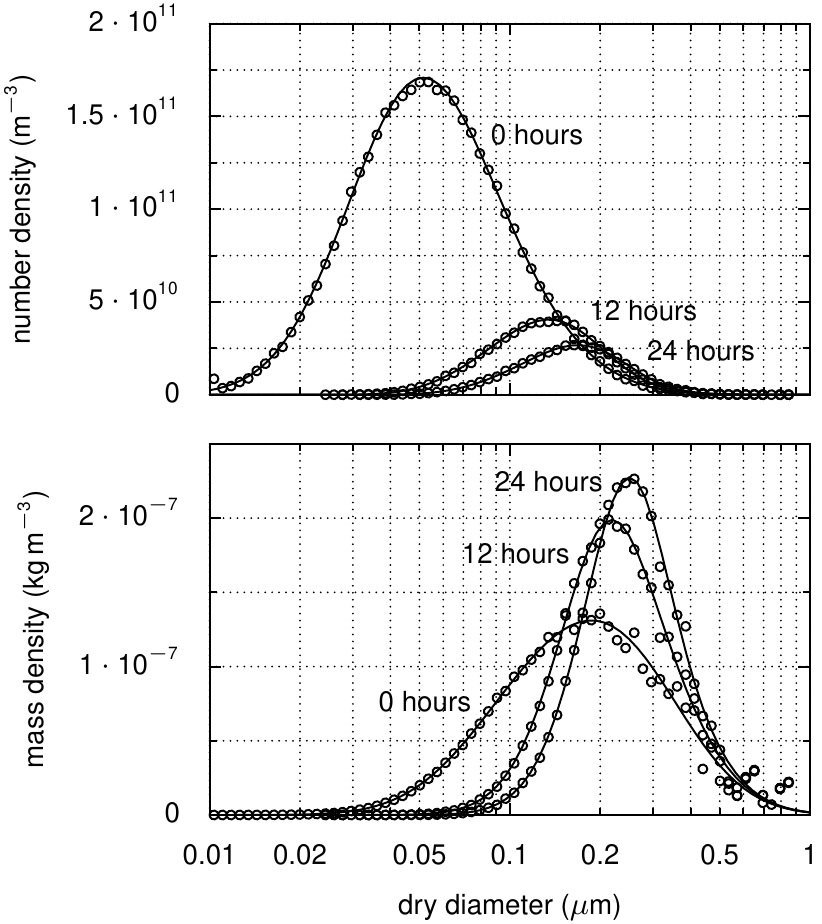}
  \end{center}
  \caption{\label{fig:test_brownian} Comparison of the stochastic
    particle-resolved method using $10^5$ particles (circles) against
    a sectional solution (lines) to the Smoluchowski equation for the
    Brownian kernel according to \citet{Jacobson1999buch}. }
\end{figure}

For verification of the PartMC stochastic coagulation method we
compared PartMC using $10^5$ particles against a sectional solution to
the Smoluchowski equation \citep{Bott1998} for a Brownian kernel
\citep{Jacobson1999buch}. For Figure~\ref{fig:test_brownian} we used
two overlapping log-normal modes as the initial condition and the
results show that we have excellent agreement for the number and mass
distributions for this test case, which is representative of the
simulation in Section~\ref{sec:urban_plume}. At the very largest sizes
there is some noise in the particle-based mass distribution, as each
individual particle has significant mass at these sizes. This noise
could be reduced by averaging several simulations in a Monte Carlo
fashion, or by using a variable number of physical particles per
computational particle, as in the Mass Flow Algorithm \citep{Ba1999,
  EiWa2001b}. We do not consider this noise to be significant enough
for the study in this paper to require amelioration.

For the Brownian kernel in Figure~\ref{fig:test_brownian} the use of
the binned stochastic coagulation algorithm of
Section~\ref{sec:partmc} improved the accept rate from 0.95\% to 86\%,
requiring more than 90 times fewer kernel evaluations. For a more
non-uniform gravitational kernel, such as found in cloud-aerosol
simulations, the binned algorithm increased the accept rate from
0.007\% to 86\%, a reduction of over 12,000 times in the number of
kernel evaluations (not shown in a figure).

\begin{figure}
  \begin{center}
    \includegraphics{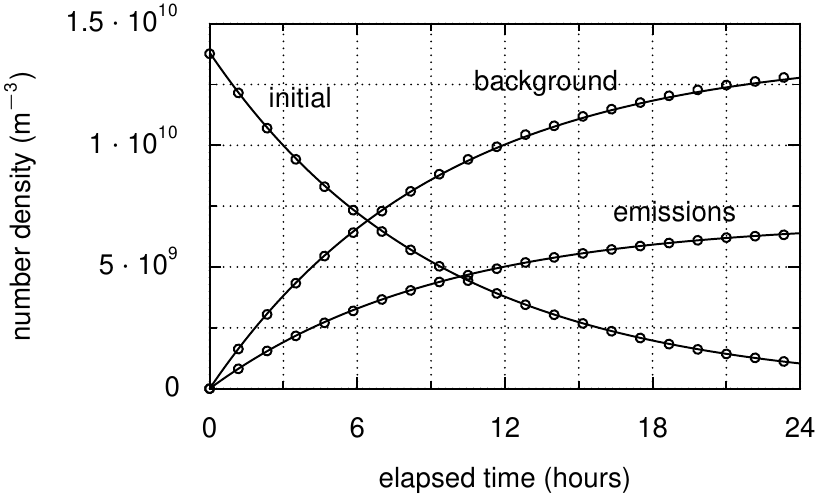}
  \end{center}
  \caption{\label{fig:test_emission} Comparison of the stochastic
    particle-resolved method using $10^5$ particles (circles) against
    the analytical solution (lines) for a simulation with only
    constant mean rate emissions and dilution.}
\end{figure}

Figure~\ref{fig:test_emission} compares the stochastic treatment of
emissions and dilution using $10^5$ particles against the analytical
solution for constant mean emission and dilution rates. This test also
shows excellent agreement. We thus see that PartMC-MOSAIC is
performing emissions, dilution, and coagulation accurately, and the
chemistry modeling is of similar accuracy to that in current
state-of-the-art sectional and modal aerosol models.

\section{Application of PartMC-MOSAIC to an idealized urban plume scenario}
\label{sec:urban_plume}

\subsection{Setup of case study}
\label{sec:case_study_setup}

For this study we considered an idealized urban plume scenario. We
tracked the evolution of gas phase species and aerosol particles in a
Lagrangian air parcel that initially contains background air and is
advected over and beyond a large urban area. The simulation started at
06:00 local standard time (LST), and during the advection process,
primary trace gases and aerosol particles from different sources were
emitted into the air parcel for 12 hours. After 18:00 LST, the
emissions were switched off, and the evolution of the air parcel was
tracked for another 12 hours.

\begin{table*}
  \begin{center}
    \begin{tabular}{c c c}
      \hline
      MOSAIC species & Concentration in ppb
      & Emissions in $\rm nmol \, m^{-2} \, s^{-1}$\\
      \hline
      NO      &     0.1               & 31.8 \\
      $\rm NO_2$     &     1.0               & 1.67 \\
      $\rm HNO_3$    &     1.0               &  - \\
      $\rm O_3$      &     50.0              &  -\\
      $\rm H_2O_2$    &     1.1               &  - \\
      CO      &     21                & 291.3 \\
      $\rm SO_2$     &     0.8               & 2.51 \\
      $\rm NH_3$     &     0.5               & 6.11 \\
      HCl     &     0.7               &  - \\
      $\rm CH_4$     &     2200              &  - \\
      $\rm C_2H_6$    &     1.0               &  - \\
      HCHO    &     1.2               & 1.68 \\
      CH3OH   &     0.12              & 0.28 \\
      CH3OOH  &     0.5               & -  \\
      ALD2    &     1.0               & 0.68 \\
      PAR     &     2.0               & 96 \\
      AONE    &     1.0               & 1.23 \\
      ETH     &     0.2               & 7.2 \\
      OLET    &     $2.3\cdot 10^{-2}$ & 2.42 \\
      OLEI    &     $3.1\cdot 10^{-4}$ & 2.42 \\
      TOL     &     0.1               & 4.04 \\
      XYL     &     0.1               & 2.41 \\
      ONIT    &     0.1               & - \\
      PAN     &     0.8               & - \\
      RCOOH   &     0.2               & - \\
      ROOH    &     $2.5\cdot 10^{-2}$ & - \\
      ISOP    &     0.5               & 0.23 \\
      ANOL    &     -                 & 3.45  \\
      \hline
    \end{tabular}
  \end{center}
  \caption{Initial concentrations of gas phase species and average gas
    phase emissions. The emissions represent area emissions and are
    averaged over the 12-hour emission period. We obtain the volume
    emission rate $\dot{\vec{g}}_{\rm emit}(t)$ in
    equation~(\ref{eqn:main_gas_emit_dilute}) by dividing by the mixing
    height $H(t)$. \label{tab:gas_dat}}
\end{table*}

\begin{table*}
  \begin{center}
    \begin{tabular}{c c c c c}
      \hline
      & $N$ in $\rm m^{-3}$               & $D_{\rm gn}$ in $\rm \mu m$
      & $\sigma_{\rm g}$
      & Composition \\
      \hline
      Initial mode 1 & $3.2\cdot 10^{9}$  & 0.02                    & 1.45
      & 50\% $\rm (NH_4)_2SO_4$, 50\% POM \\
      Initial mode 2 & $2.9\cdot 10^{9}$  & 0.116                   & 1.65
      & 50\% $\rm (NH_4)_2SO_4$, 50\% POM\\
      \hline
      & $E$ in $\rm m^{-2}\, s^{-1}$       & $D_{\rm gn}$ in $\rm \mu m$
      & $\sigma_{\rm g}$
      & Composition\\
      \hline
      Meat cooking & $9\cdot 10^{6}$      & 0.086                   & 1.9
      & POM \\
      Diesel vehicles & $1.6\cdot 10^{8}$ & 0.05                    & 1.7
      & 10\% POM, 90\% BC \\
      Gasoline vehicles& $5\cdot 10^{7}$  & 0.05                    & 1.7
      & 66\% POM, 34\% BC
    \end{tabular}
  \end{center}
  \caption{Initial and emitted aerosol distribution parameters, as in
    equation~(\ref{eqn:log_normal_dist}). The initial aerosol
    distribution is also used as the background aerosol
    distribution. $E$ is the area source strength of particle
    emissions. Dividing $E$ by the mixing height $H(t)$ and
    multiplying by a normalized composition distribution gives the
    number distribution emission rate $\dot{n}_{\rm emit}(\vec{\mu}, t)$ in
    equation~(\ref{eqn:main_aero_emit}). \label{tab:aero_dat}}
\end{table*}

Initial gas-phase concentrations and emissions were adapted from the
Southern California Air Quality Study (SCAQS) simulation (August
26-29, 1988 period) in \citet{Zaveri2008}, and are listed in
Table~\ref{tab:gas_dat}. Note that while gas phase emissions in the
simulation varied with time, Table~\ref{tab:gas_dat} gives only the
average over the emission period. The initial particle size
distribution, which was identical to the background aerosol
distribution, was bimodal with Aitken and accumulation modes
\citep{Jaenicke1993}. We assumed that it consisted of $\rm
(NH_4)_2SO_4$ and primary organic mass (see
Table~\ref{tab:aero_dat}). We considered three different types of
carbonaceous aerosol emissions: 1) meat cooking aerosol, 2) diesel
vehicle soot, and 3) gasoline vehicle soot. The parameters for the
size distributions of these three emission categories were based on
\citet{Eldering1996}, \citet{Kittelson2006-1}, and
\citet{Kittelson2006-2}, respectively. The emission rates and the
compositions were adapted from the California Air Resources Board
database \citep{CARBemissions2008}.

For simplicity in this idealized study, the particle emissions
strength and their size and composition were kept constant with
time. Furthermore, we assumed that the particles from these sources
were emitted as fully-internal mixtures of the species listed in
Table~\ref{tab:aero_dat}, since to date the mixing state of particle
emissions is still not well characterized. Hence, it is difficult to
justify a more sophisticated treatment for the emissions. However,
once this data is available it will be straightforward to implement
the information accordingly in our model. Sea salt, biomass burning
and mineral dust particles as well as particles from biological
sources (e.g. pollen) were not treated in this test case.

\begin{figure*}
  \begin{center}
    \includegraphics{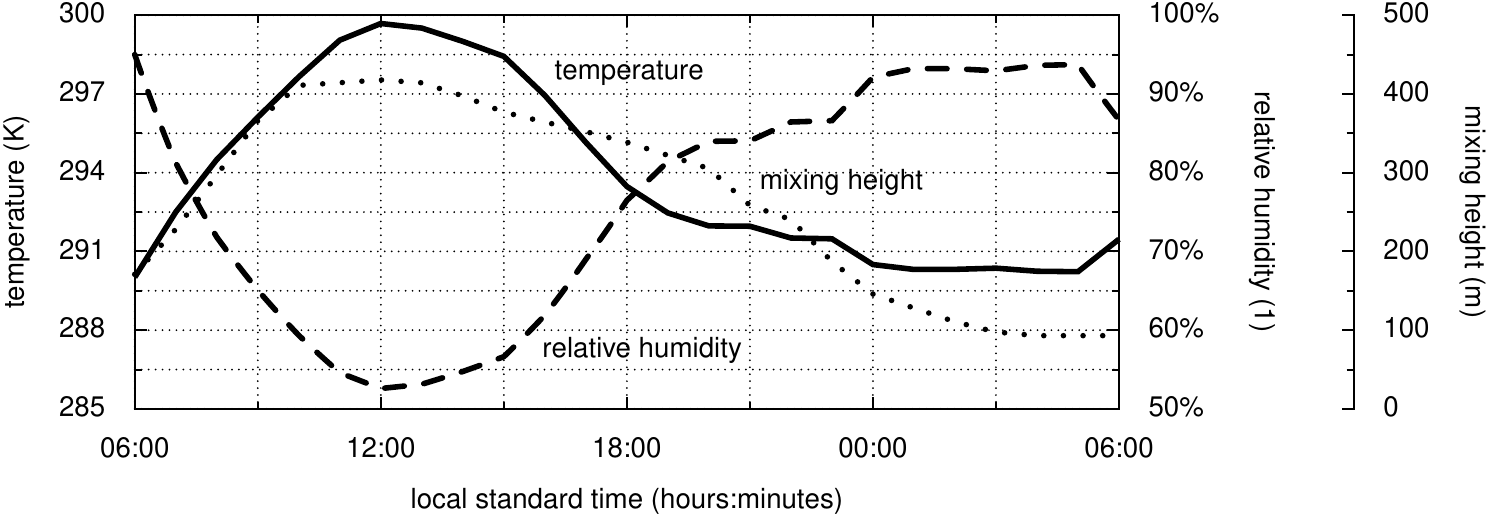}
  \end{center}
  \caption{\label{fig:env} Time series of temperature,
    relative humidity, and mixing height over the course of the
    24-hour simulation.}
\end{figure*}

Before we discuss the results on aerosol mixing state in detail we
provide the context for the conditions in our case study with
Figures~\ref{fig:env} to \ref{fig:aero_num_dist}. We did not attempt
to simulate a specific episode or trajectory for the Los Angeles basin
(as was done in \citet{Kleeman1997}), but rather an idealized urban
plume scenario, with conditions that were consistent with a polluted
environment. The temperature, relative humidity, and mixing height
along the trajectory were adapted from spatially-averaged values from
the Los Angeles Air Basin simulation of \citet{Zaveri2008} and
references therein. The temperature and mixing height were prescribed
as functions of time, while the pressure and water mixing ratio were
kept constant and the relative humidity and dry density were updated
accordingly. The variation of these parameters is shown in
Figure~\ref{fig:env}. The relative humidity started at 95\%, then
decreased to 53\% during the day and increased again to 94\% during
the following night. As we show below, the diurnal cycle of the
ambient conditions impacted the thermodynamic equilibria and the phase
states of the particles.

An increase of the mixing height during the morning caused dilution of
the gas and aerosol concentrations within the air parcel and was
accompanied by entrainment of background air, as discussed in
Section~\ref{sec:emission_dilution}. We also considered dilution due
to horizontal turbulent diffusion, using a first-order dilution rate
of $1.5 \cdot 10^{-5}\rm m \,s^{-1}$.

We resolved the total aerosol distribution with $10^5$ Monte Carlo
particles initially. The corresponding initial total number density
was $N=6.1 \cdot 10^9 \rm \, m^{-3}$ and so the computational volume
was initially $V = N_{\rm MC}/N = 16\rm\,cm^3$. It remained between $V
= 8\rm\,cm^3$ and $V = 17\rm\,cm^3$ for the duration of the run as
particle number $N_{\rm MC}$ and number density $N$ changed due to
emissions, dilution, and coagulation. The number of particles remained
between $N_{\rm MC} = 199799$ and $N_{\rm MC} = 60655$. The timestep
used for this simulation was $\Delta t = 1\rm\,minute$. While better
estimates of the system statistics could be obtained with multiple
realizations, we found a single run to give reasonable results in this
case, as demonstrated in Figures~\ref{fig:test_brownian} and
\ref{fig:test_emission} and discussed in
Section~\ref{sec:verification}. Although not shown here, runs with
different random initialization gave essentially the same results.

To quantify the impact of coagulation we performed two runs, one base
case including coagulation as described above, and one case without
coagulation. Otherwise, the conditions for the two runs were
identical.

\subsection{Gas species evolution}

\begin{figure}
  \begin{center}
    \includegraphics{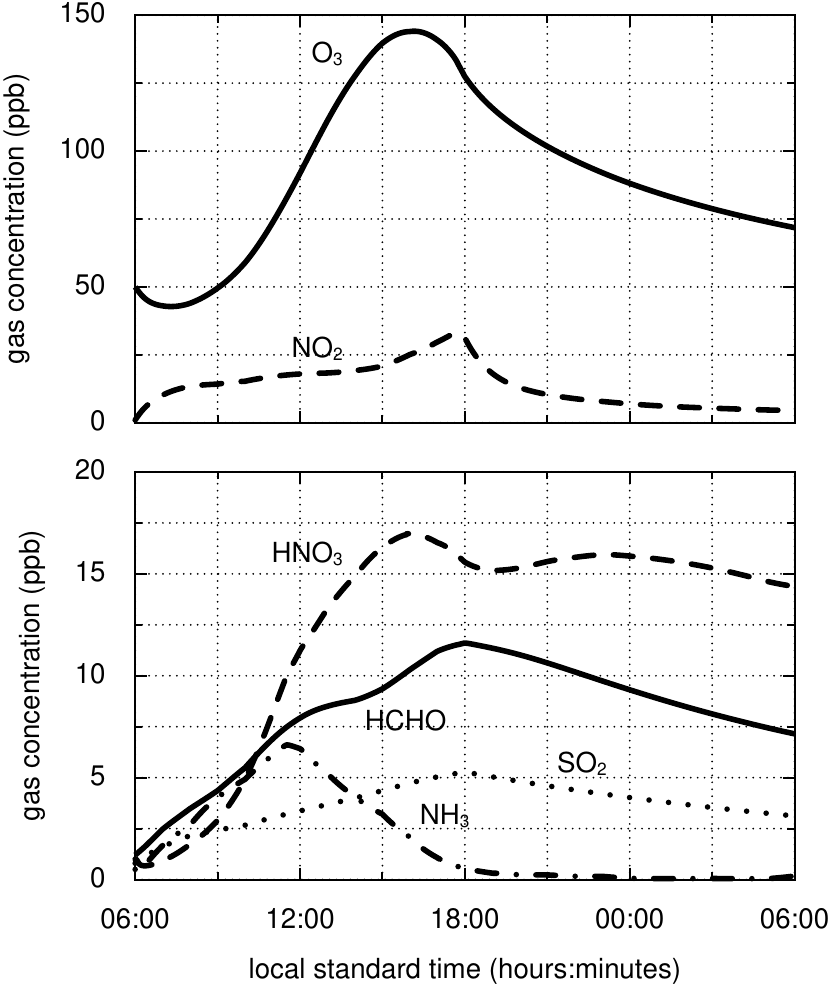}
  \end{center}
  \caption{\label{fig:gas} Time series of selected gas phase species.}
\end{figure}

Figure~\ref{fig:gas} shows the evolution of selected gas phase species
undergoing a diurnal cycle typical for a photochemistry episode under
polluted conditions. During the daytime we observed a considerable
production of $\rm O_3$, reaching a maximum value of 144~ppb at 16:09
LST. The $\rm NO_2$ concentration increased up to 33~ppb during the
time that $\rm NO_x$ was emitted, and decreased after 18:00 LST due to
dilution and chemical reactions after the emissions had stopped. $\rm
HNO_3$ reached 17~ppb and contributed to the formation of ammonium
nitrate in the particle phase. $\rm NH_3$ levels reached 6.6~ppb
during the daytime and later vanished due to gas-to-particle
conversion. HCHO was both emitted and chemically produced with a
maximum value of 12~ppb at 17:59 LST.

\subsection{Bulk aerosol evolution}

\begin{figure}
  \begin{center}
    \includegraphics{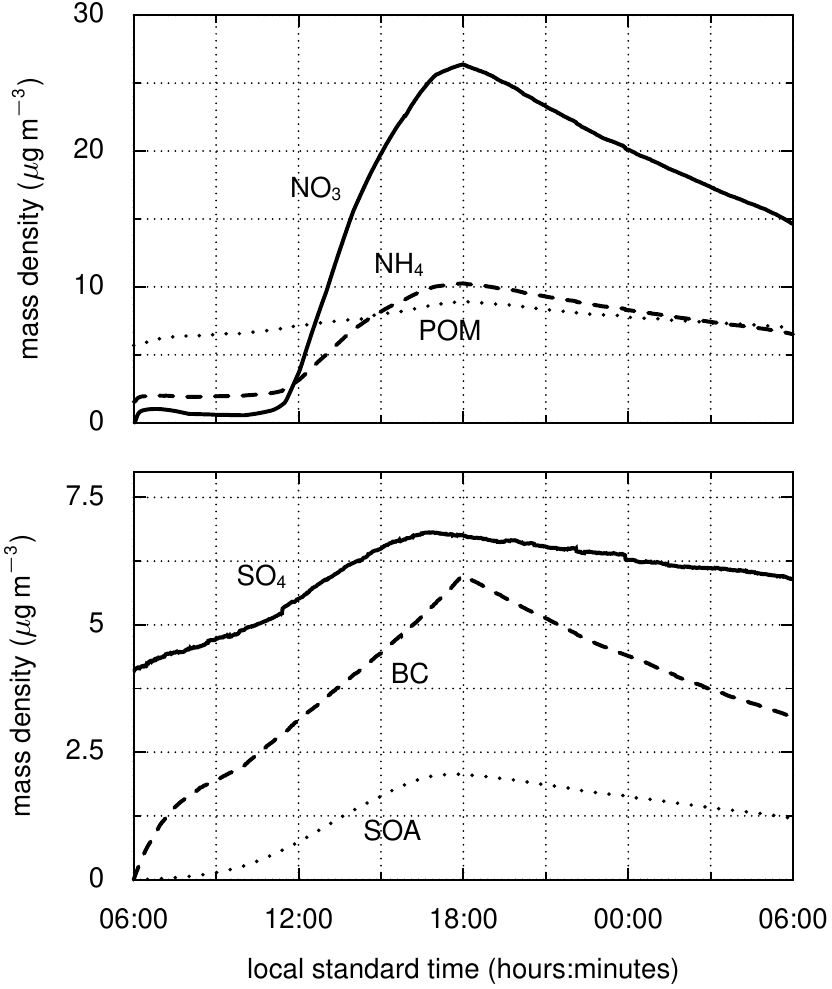}
  \end{center}
  \caption{\label{fig:aero_time} Time series of total mass densities
    of selected aerosol species: $M_{\rm NO_3}$, $M_{\rm NH_4}$,
    $M_{\rm POM}$, $M_{\rm SO_4}$, $M_{\rm BC}$, and $M_{\rm SOA}$.}
\end{figure}

Figure~\ref{fig:aero_time} shows time series of the bulk aerosol
concentrations. We observe a pronounced production of ammonium
nitrate, reaching nitrate concentrations of up to $26\rm \, \mu
g\,m^{-3}$ and ammonium concentration of $10.3\rm \, \mu g\,m^{-3}$ in
the late afternoon. Sulfate concentrations increased from $4.1\rm \,
\mu g\,m^{-3}$ to $6.8\rm \, \mu g\,m^{-3}$ due to condensation of
photochemically produced sulfuric acid. POM and BC were directly
emitted and accumulated to $9.0\rm \, \mu g\,m^{-3}$ and $6.0\rm \,
\mu g\,m^{-3}$, respectively, until 18:00 LST when the emissions
stopped.  After 18:00 LST the mass densities declined due to dilution,
especially nitrate and BC for which the background concentration was
zero.

\subsection{Aerosol distribution functions} 
\label{sec:aer_dist_func}

We take $N(D)$ ($\rm m^{-3}$) to be the cumulative number
distribution, giving the number of particles per volume that have
diameter less than $D$. Similarly, the cumulative mass distribution
$M(D)$ ($\rm kg\,m^{-3}$) gives the mass per volume of particles with
diameter less than $D$, while the per-species cumulative mass
distribution $M_x(D)$ gives the mass per volume of species $x$ in
particles with diameter less than $D$. We write $N = N(\infty)$, $M =
M(\infty)$, and $M_x = M_x(\infty)$ for the number, mass, and
per-species mass densities, respectively.

Given the cumulative densities, we define the number distribution
$n(D)$ ($\rm m^{-3}$), mass distribution $m(D)$ ($\rm kg\,m^{-3}$) and
per-species mass distribution $m_x(D)$ ($\rm kg\,m^{-3}$) by
\begin{align}
n(D) &= \frac{d N(D)}{d \log_{10} D} \\
m(D) &= \frac{d M(D)}{d \log_{10} D} \\
m_x(D) &= \frac{d M_x(D)}{d \log_{10} D}.
\end{align}
The initial, background, and emitted number densities used in this
paper will all be superpositions of log-normal distributions, each
defined by
\begin{equation}
  \label{eqn:log_normal_dist}
  n(D) = \frac{N}{\sqrt{2\pi} \log_{10} \sigma_{\rm g}} \exp\left(
  - \frac{(\log_{10} D - \log_{10} D_{\rm gn})^2}{2 (\log_{10} \sigma_{\rm g})^2}
  \right)
\end{equation}
where $N$~($\rm m^{-3}$) is the total number density, $D_{\rm
  gn}$~($\rm m$) is the geometric mean diameter, and $\sigma_{\rm g}$
(dimensionless) is the geometric standard deviation.

To discuss the composition of a particle, we refer to certain mass
fractions of species, as
\begin{align}
f_{\rm BC,POM} &= \frac{\mu_{\rm BC}}{\mu_{\rm BC} + \mu_{\rm POM}} \\
f_{\rm BC,dry} &= \frac{\mu_{\rm BC}}{\mu_{\rm dry}} \\
f_{\rm H_2O,all} &= \frac{\mu_{\rm H_2O}}{\mu_{\rm all}},
\end{align}
where we recall that $\mu_x$ ($\rm m^3$) is the mass of species $x$ in
a given particle, $\mu_{\rm all}$ ($\rm m^3$) is the total wet mass of
the particle, and $\mu_{\rm dry} = \mu_{\rm all} - \mu_{\rm H_2O}$
($\rm m^3$) is the total dry mass.

We extend the number and mass densities to be functions of both
particle composition and diameter. That is, the two-dimensional
cumulative number distribution $N_{y,x}(f, D)$ ($\rm m^{-3}$) is the
number of particles per volume that have a mass ratio of $y$ to $x$
less than $f$ and a diameter less than $D$. The two-dimensional number
distribution $n_{y,x}(f, D)$ ($\rm m^{-3}$) is then defined by
\begin{equation}
\label{eqn:N_f_D_defn}
 n_{y,x}(f, D)
= \frac{\partial^2 N_{y,x}(f,D)}{\partial f
\, \partial \log_{10} D}.
\end{equation}
The two-dimensional mass distribution $m_{y,x}(f,D)$ ($\rm
kg\,m^{-3}$) and cumulative mass distribution $M_{y,x}(f, D)$ ($\rm
kg\,m^{-3}$) are defined similarly.

We can also define two-dimensional densities based on other particle
quantities. In particular, if we denote by $k$ the number of
coagulation events that a given particle has experienced during the
simulation time then we can define the two-dimensional
singly-cumulative number distribution $N_{\rm coag}(k, D)$ ($\rm
m^{-3}$) to be the number of particles per volume with $k$ coagulation
events and diameter less than $D$. Then $n_{\rm coag}(k, D)$ ($\rm
m^{-3}$) is defined by
\begin{equation}
\label{eqn:n_coag}
  n_{\rm coag}(k, D) = \frac{\partial N_{\rm coag}(k, D)}{\partial
    \log_{10} D}.
\end{equation}

For ease of comparison between different plots we frequently use
normalized densities denoted by a hat, so the normalized
two-dimensional number distribution $\hat{n}_{y,x}(f, D)$
(dimensionless) is defined by
\begin{equation}
  \hat{n}_{y,x}(f, D) = \frac{n_{y,x}(f, D)}{N},
\end{equation}
and similarly for the mass distribution.

We also find it convenient to plot one-dimensional mass densities for
certain composition ranges, as done in
Figure~\ref{fig:aero_bc_mixing}.  We write $m_{\rm BC,dry}([f_1, f_2],
D)$ ($\rm kg\,m^{-3}$) to refer to the total mass distribution
(including water), where $f_{\rm BC,dry}$ is between $f_1$ and $f_2$,
\begin{equation}
\label{eqn:mass_den_range}
  m_{\rm BC,dry}([f_1,f_2], D) = M_{\rm BC,dry}(f_2, D) - M_{\rm BC,dry}(f_1, D).
\end{equation}

\subsection{Aerosol size distribution evolution}
\label{eqn:aero_size_dist_evolution}

\begin{figure}
  \begin{center}
    \includegraphics{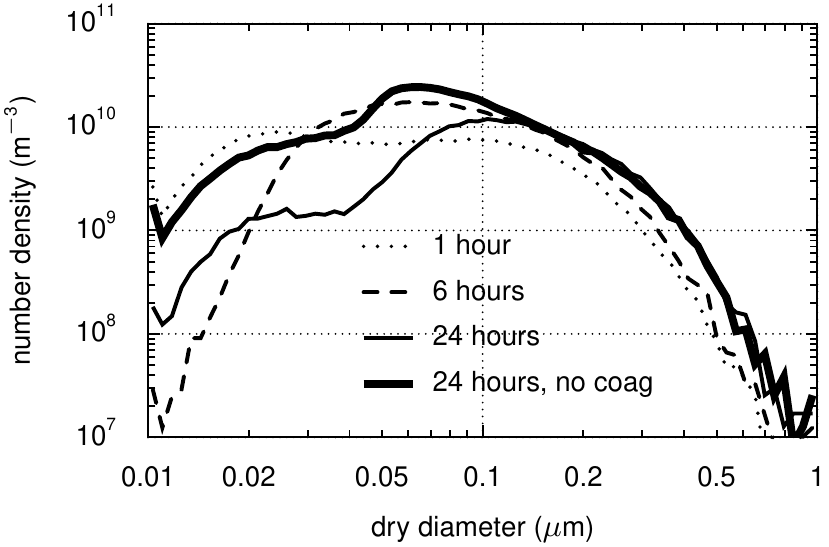}
  \end{center}
  \caption{\label{fig:aero_num_dist} Number size distributions $n(D)$
    for the simulation with coagulation after 1, 6, and 24 hours. For
    comparison the distribution without coagulation after 24 hours is
    also shown.}
\end{figure}

Figure~\ref{fig:aero_num_dist} shows results of the number size
distributions $n(D)$ after 1, 6, and 24 hours of simulation including
coagulation. For comparison, the result after 24 hours of simulation
without coagulation is also shown. The distribution after 1 hour still
resembled the bimodal initial distribution (compare
Table~\ref{tab:aero_dat}), which was identical to the background
distribution. After 6 hours the distribution was primarily determined
by the emissions. Concurrently, condensation of secondary species
caused aerosol growth. Particles at small sizes were depleted due to
coagulation. After 24 hours the Aitken mode of the background appeared
again as a result of dilution. Compared to the size distribution
without coagulation, the size distribution with coagulation showed a
substantial decrease in number density for particles smaller than
0.1~$\rm \mu m$. With coagulation the total number density $N$ peaked
at the end of the emission period after 12 hours with a maximal value
of $1.67 \cdot 10^{10} \, \rm m^{-3}$. After this, $N$ declined due to
coagulation and dilution to $7.36\cdot 10^{9} \rm \, m^{-3}$. The
simulation without coagulation lead to a maximum total number density
of $2.39\cdot 10^{10} \, \rm m^{-3}$, and a final value of $1.54\cdot
10^{10} \, \rm m^{-3}$. This means that coagulation decreased the peak
and final total number concentrations by 30\% and 52\%,
respectively. Comparing the number densities for the specific
diameters $D = 0.03$, $0.05$, $0.07$, and $0.1\rm \, \mu m$ with and
without coagulation, we find that coagulation decreased the number
density $n(D)$ by 82\%, 84\%, 66\%, and 29\% respectively.

We notice that for all size distributions shown in
Figure~\ref{fig:aero_num_dist} the results are somewhat ``noisy'' at
small and large diameters. This noise is inherent to the stochastic
model that is used for coagulation, dilution and emissions. Towards
the edges of the size spectrum only a few particles are being used to
represent the size distribution due to the low number density. Single
particle variations arising from the stochastic model thus appear as a
noisy curve. This could be rectified by averaging repeated Monte Carlo
simulations or by using a variable number of physical particles per
computational particle, as in the Mass Flow Algorithm \citep{Ba1999,
  EiWa2001b}.

\subsection{Aerosol mixing state evolution}
\label{eqn:aero_mix_state_evolution}

\begin{figure*}
  \begin{center}
    \includegraphics{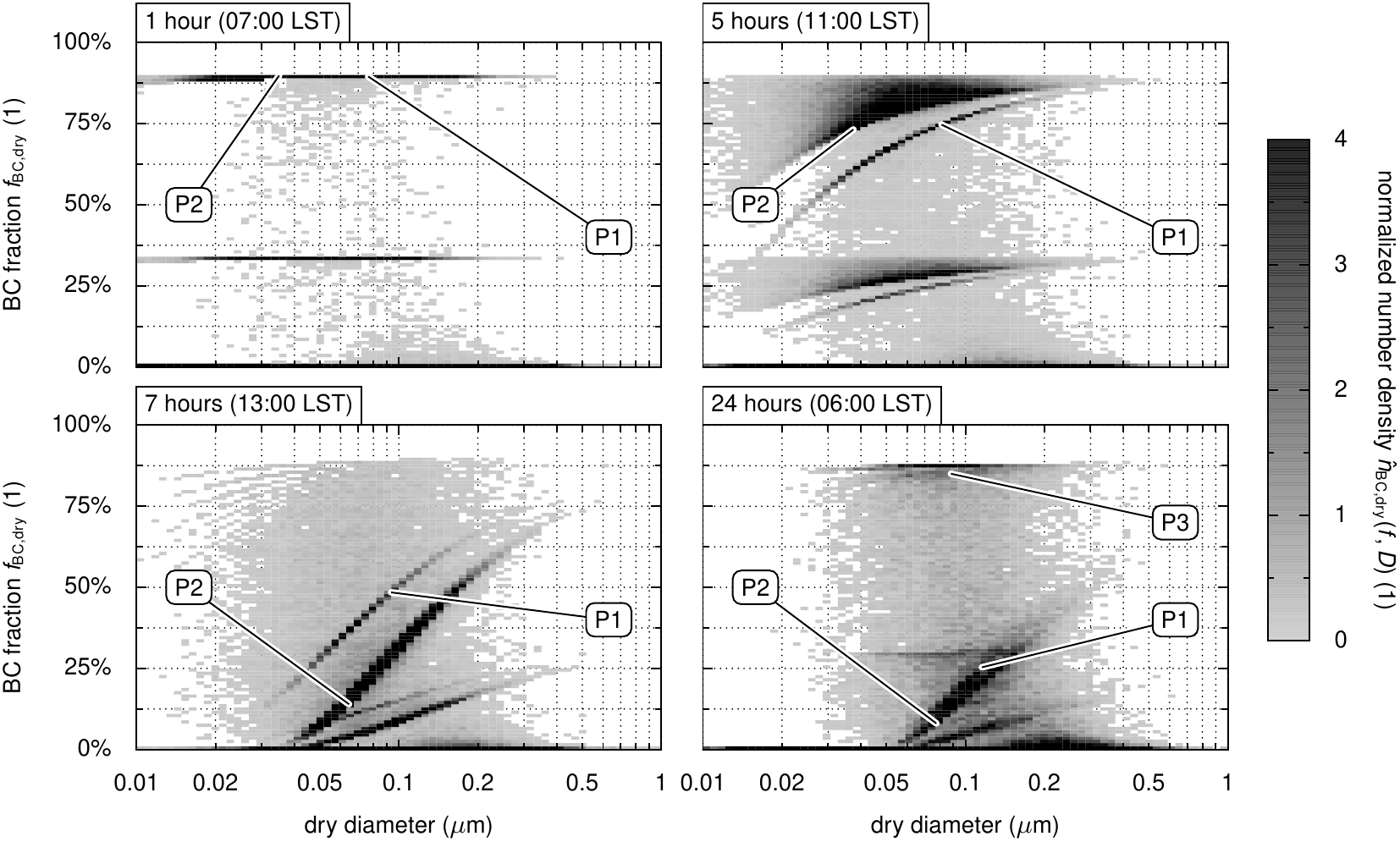}
  \end{center}
  \caption{\label{fig:aero_2d_all} Normalized two-dimensional number
    distribution $\hat{n}_{{\rm BC},{\rm dry}}(f, D)$ after 1, 5, 7, and 24
    hours of simulation. The labels P1, P2, and P3 track three
    individual diesel emission particles as they evolve over the
    course of the simulation. The maximum plotted value for
    $\hat{n}_{{\rm BC},{\rm dry}}(f, D)$ is capped at 4 to allow
    better resolution. }
\end{figure*}

While Figures~\ref{fig:aero_time} and \ref{fig:aero_num_dist} give an
overview of aerosol size distribution and composition just like we
obtain from traditional distribution-based models, they do not address
the issue of mixing state. To elucidate how the mixing state evolved
over the course of the simulation we display the data as shown in
Figure~\ref{fig:aero_2d_all}. The panels show the two-dimensional
number distributions as a function of dry size and mass fraction of
BC, $f_{{\rm BC},{\rm dry}}$, after 1, 5, 7, and 24 hours of
simulation. This corresponds to LST 07:00, 11:00, 13:00, and 06:00 of
the next morning. Our definition of the two-dimensional size
distribution is described in Section \ref{sec:aer_dist_func} above.

\begin{figure*}
  \begin{center}
    \includegraphics{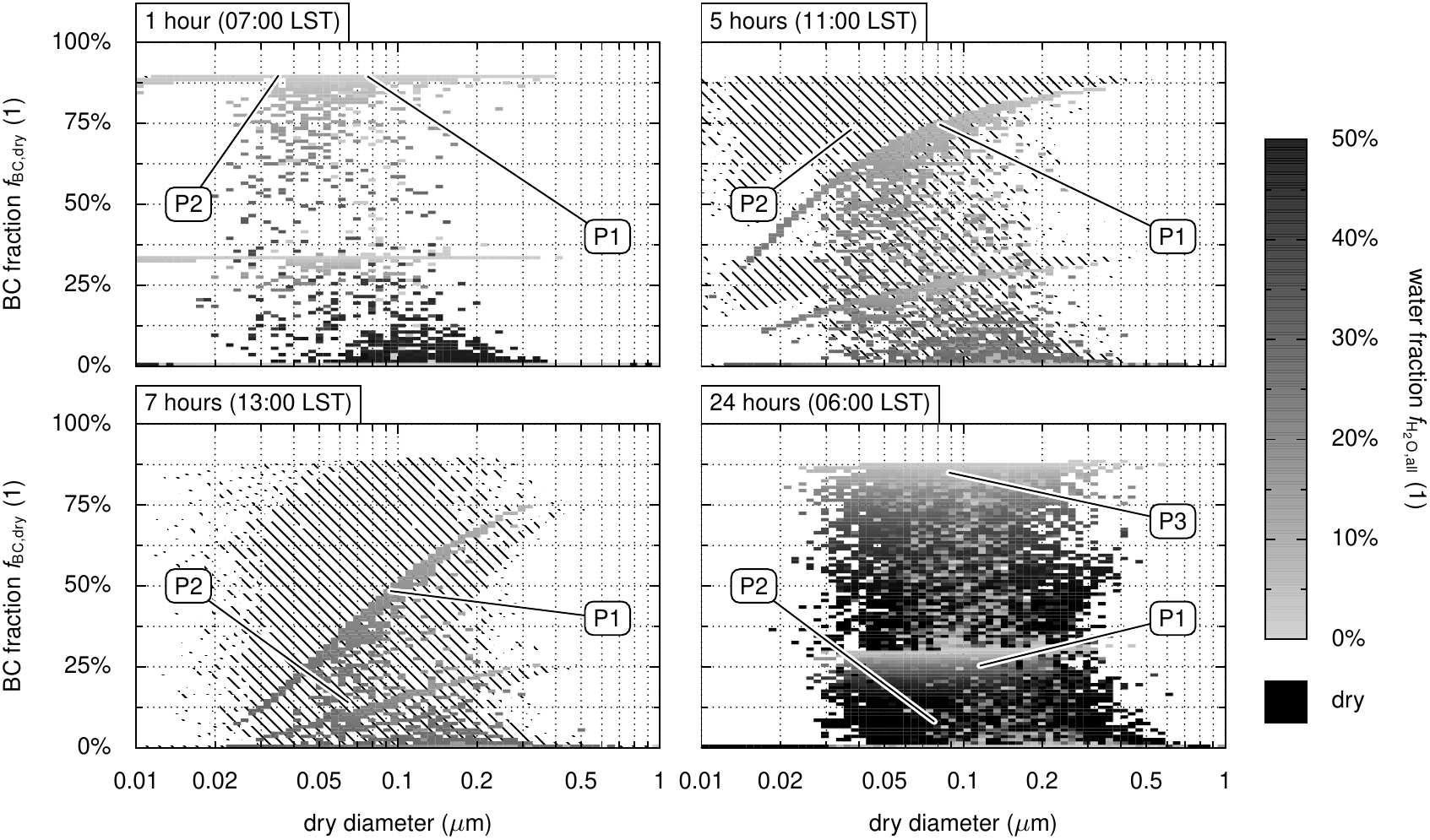}
  \end{center}
  \caption{\label{fig:aero_2d_water} Water fraction, $f_{\rm H_2O,
      all}$ as a function of BC fraction $f_{\rm BC, dry}$ and dry
    diameter after 1, 5, 7, and 24 hours of simulation. The labels P1,
    P2, and P3 track three individual diesel emission particles as
    they evolve over the course of the simulation. Note that the water
    fraction of wet particles is plotted over the hashing for dry
    particles and sometimes obscures it. In particular, after 1 hour
    (07:00 LST) there are dry diesel and gasoline particles present
    but they are not visible. The water fraction plotted for a given
    two-dimensional bin is the minimum of the water fraction for all
    wet particles in that bin. For example, after 24 hours the
    particle P1 is very wet (see Figure~\ref{fig:aero_particles}) but
    there are much drier particles present with similar composition,
    giving a low $f_{\rm H_2O, all}$ value on the plot at P1. The
    maximum plotted value for $f_{\rm H_2O, all}$ is capped at 50\% to
    allow better resolution.}
\end{figure*}

\begin{figure}
  \begin{center}
    \includegraphics{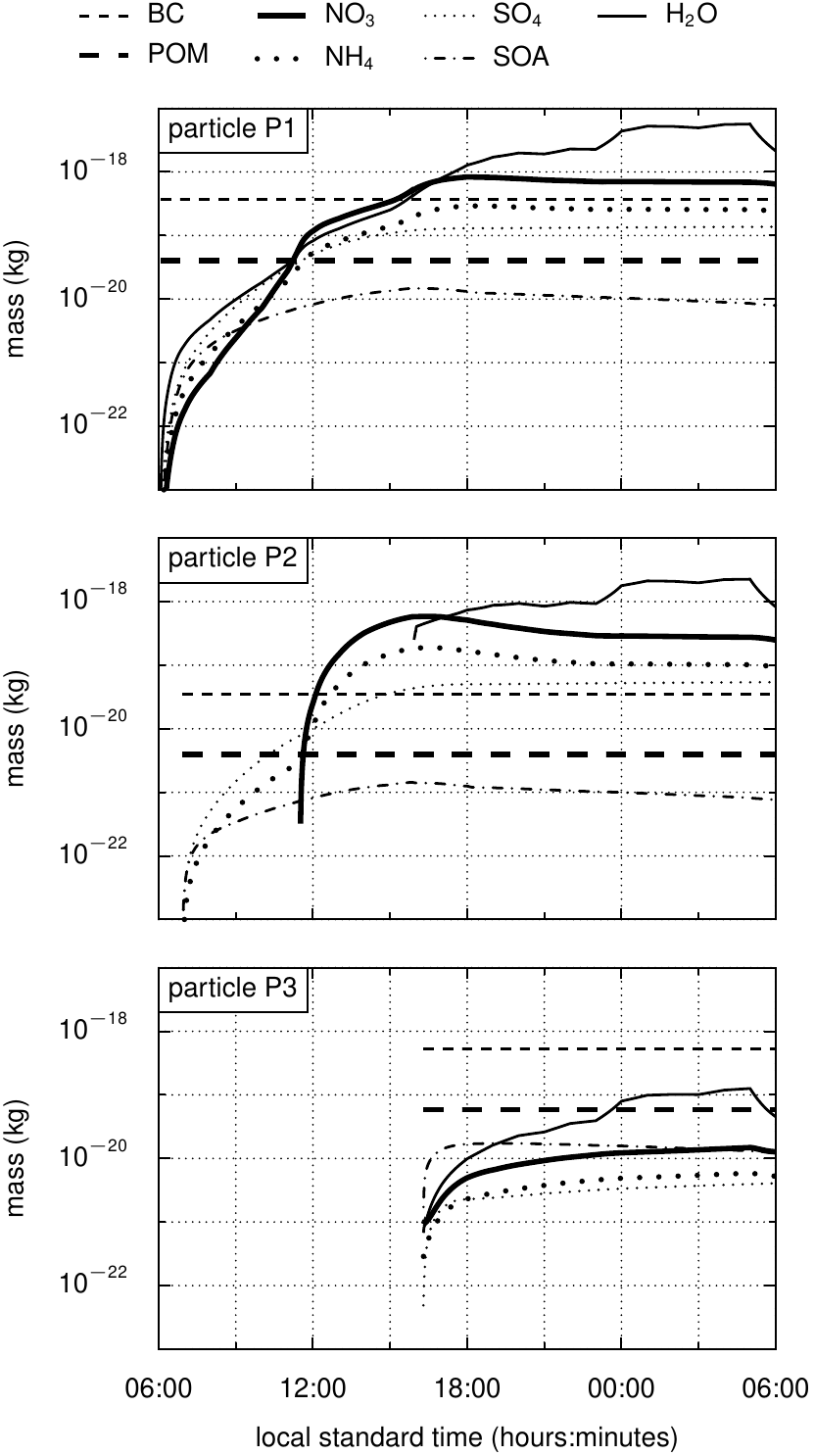}
  \end{center}
  \caption{\label{fig:aero_particles} Time history of the composition
    of three individual diesel particles, P1, P2, and P3. P1 is
    emitted at 06:05 LST and always contains water. P2 is emitted at
    06:56 LST and is initially dry but becomes wet in the
    afternoon. P3 is emitted at 16:18 LST later in the day when little
    condensation occurs.}
\end{figure}

We will discuss the evolution of the two-dimensional number size
distribution in conjunction with Figure~\ref{fig:aero_2d_water}. The
gray scale in this figure shows the water content of the particles,
$f_{{\rm H_2O},{\rm all}}$, as a function of BC mixing state, $f_{{\rm
    BC},{\rm dry}}$, and particle size. We also include in
Figure~\ref{fig:aero_particles} the temporal evolution of the
composition of three representative particles to aid the
interpretation. These particles are labeled with P1, P2 and P3 in
Figure~\ref{fig:aero_2d_all}.

Figure~\ref{fig:aero_2d_all} shows the BC mixing state, $f_{{\rm
    BC},{\rm dry}}$, relative to all other dry constituents. Since
even at the time of emission no particles were pure BC, particles were
not present at $f_{{\rm BC},{\rm dry}} = 100\%$. Fresh emissions from
diesel vehicles ($f_{{\rm BC},{\rm dry}} = 90\%$) and gasoline
vehicles ($f_{{\rm BC},{\rm dry}} = 34\%$) appear as horizontal lines
since particles in one emission category were all emitted with the
same composition. At $f_{{\rm BC},{\rm dry}} = 0\%$ all the particles
appear that do not contain any BC (i.e. background particles and
particles from meat cooking emissions that have not undergone
coagulation with particles containing BC). After 1 hour (07:00 LST) a
small number of particles in between these three classes indicate the
occurrence of coagulation.

Under the initial ambient conditions the emitted diesel and gasoline
particles accumulated small amounts of ammonium sulfate, ammonium
nitrate and water. After 06:44 LST the relative humidity fell below
85\%, which is the deliquescence point of the inorganic mixture of
ammonium, sulfate and nitrate. As a result of the hysteresis of
particle deliquescence and crystallization, the particles that had
been emitted up to this point stayed wet throughout the whole day
(since the relative humidity never fell below the crystallization
point), but freshly emitted particles were dry from this point in time
onwards until the relative humidity reached 61\% in the following
afternoon at 16:00 LST. Hence, between 06:46 and 16:00 LST, wet and
dry particles co-existed in the air parcel. Particle P1 in
Figures~\ref{fig:aero_2d_all} and \ref{fig:aero_2d_water} is one of
the particles that was emitted early and stayed wet throughout the
simulation, whereas particle P2 started out dry and became wet only in
the afternoon. For the wet and dry particles different thermodynamic
equilibria applied which was reflected in the different development of
their $f_{\rm BC,dry}$ values.

As the single-particle plot Figure~\ref{fig:aero_particles} shows, the
wet particles contained nitrate from the beginning and kept taking up
nitrate, while during the first few hours vapor pressures of $\rm
HNO_3$ and $\rm NH_3$ were too low to allow nitrate formation on dry
particles. Due to this difference in nitrate formation, after 5 hours
(11:00 LST) the wet particles appear distinct from the dry particles
in Figure~\ref{fig:aero_2d_all} and reached lower $f_{\rm BC,dry}$
values, reflecting their larger ammonium nitrate content.

This changed after 11:30 LST. At this time $\rm HNO_3$ and $\rm NH_3$
were high enough so that ammonium nitrate formed on the dry
particles. They accumulated ammonium nitrate quickly, and $f_{\rm
  BC,dry}$ decreased rapidly for the dry particles. As a result,
$f_{\rm BC,dry}$ of the dry particles fell below $f_{\rm BC,dry}$ of
the wet particles, as is evident in the graph for 7 hours (13:00 LST)
in Figures~\ref{fig:aero_2d_all} and \ref{fig:aero_2d_water}.

After 18:00 LST the ammonium nitrate formations stopped, as $\rm NH_3$
concentration dropped to near zero (compare
Figure~\ref{fig:gas}). Therefore the fresh particle emissions after
this time did not accumulate much condensable material and stayed at
high $f_{\rm BC,dry}$ values. This is reflected in the single-particle
plot Figure~\ref{fig:aero_particles}, which shows the diesel particle
P3 that was emitted in the afternoon. The mass of secondary species
for this particle was much lower than its BC content. After 12 hours
(18:00 LST) both particle and gas emissions stopped, and the particle
distribution changed mainly due to coagulation and dilution. The
particle number decreased as a result of coagulation and continued
dilution with the background, but this effect is not visible in the
normalized number densities. During the evening hours the relative
humidity increased again and particles took up a substantial amount of
water. As the 24 hour plot in Figure~\ref{fig:aero_2d_water} shows,
the water content depended on the mixing state, i.e. for a given size
we find particles with water mass-fractions between near 0\% and 66\%.

Comparing the result for the end of the simulation to the results at
previous times, we note that at the end of the simulation particles
below $D = 0.03 \rm \, \mu m$ were depleted due to coagulation. A
continuum of mixing states formed in between the extreme mixing states
of $f_{{\rm BC},{\rm dry}} = 0\%$ and $f_{{\rm BC},{\rm dry}} = 90\%$.

\begin{figure*}
  \begin{center}
    \includegraphics{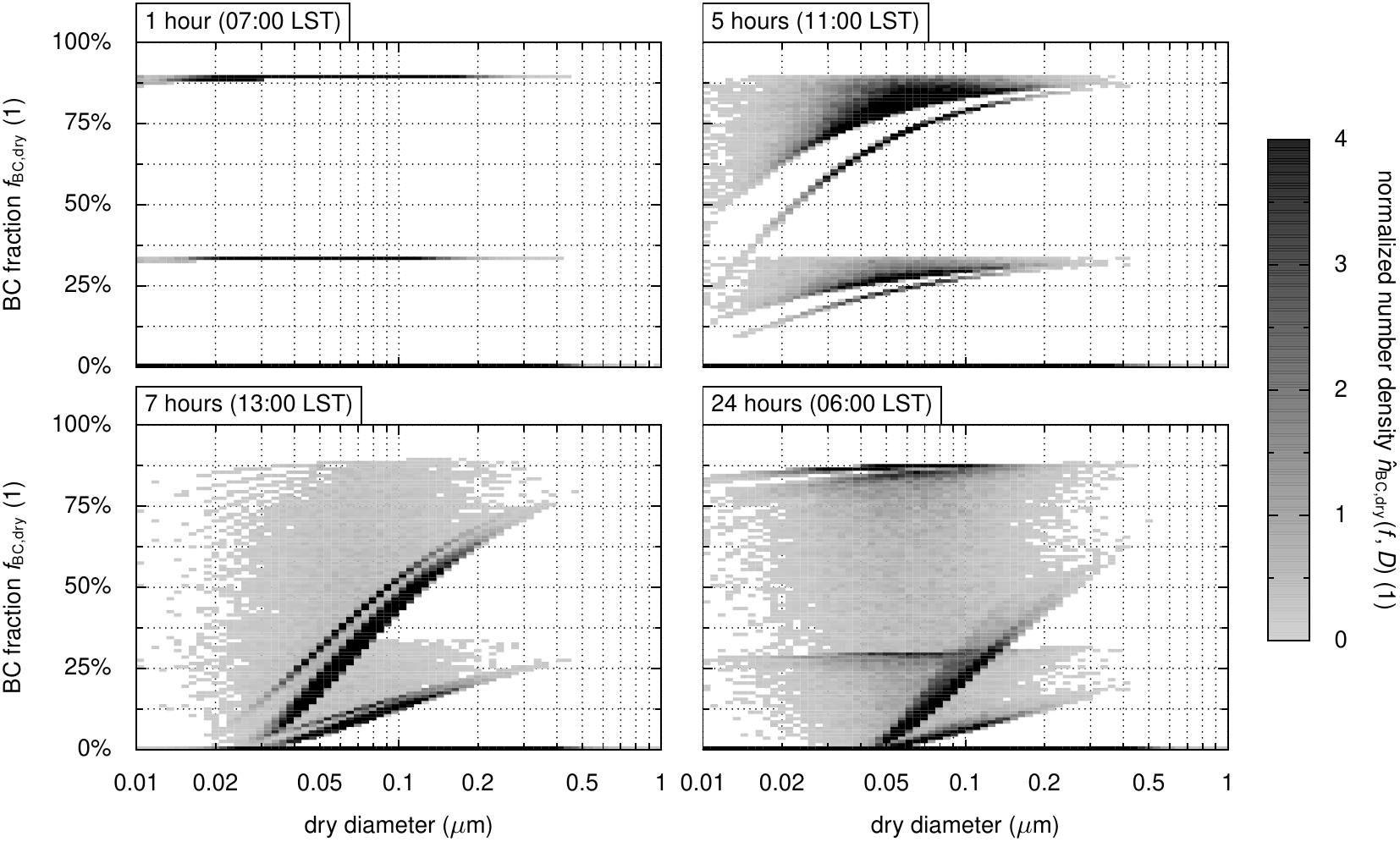}
  \end{center}
  \caption{\label{fig:aero_2d_all_no_coag} Normalized two-dimensional
    number distribution $\hat{n}_{{\rm BC},{\rm dry}}(f, D)$ after 1, 5, 7,
    and 24 hours of simulation, as in Figure~\ref{fig:aero_2d_all},
    but for the simulation without coagulation. The maximum plotted
    value for $\hat{n}_{{\rm BC},{\rm dry}}(f, D)$ is capped at 4 to
    allow better resolution.}
\end{figure*}

The comparison to the case without coagulation gives results as
displayed in Figure~\ref{fig:aero_2d_all_no_coag}. This figure is
analogous to Figure~\ref{fig:aero_2d_all} and shows the mixing state
$f_{\rm BC,dry}$ of BC with respect to the sum of all other
substances. Without coagulation similar frontal features appeared, but
diesel particles and gasoline particles remained more clearly distinct
until 6 hours of simulation (12:00 LST) without the mixing effect of
coagulation. After about 12:00 LST the mixing state became continuous,
because the most aged diesel car emissions started overlapping with
the relatively fresh gasoline emission particles. However, since these
mixing states were formed due to condensation only, an internal
mixture of primary species such as POM and BC existed only when the
particles were emitted as this mixture. After 24 hours of simulation
without coagulation, mixed particles smaller than $D = 0.03 \rm \, \mu
m$ were still present while they were depleted in
Figure~\ref{fig:aero_2d_all} with coagulation.

\begin{figure*}
  \begin{center}
    \includegraphics{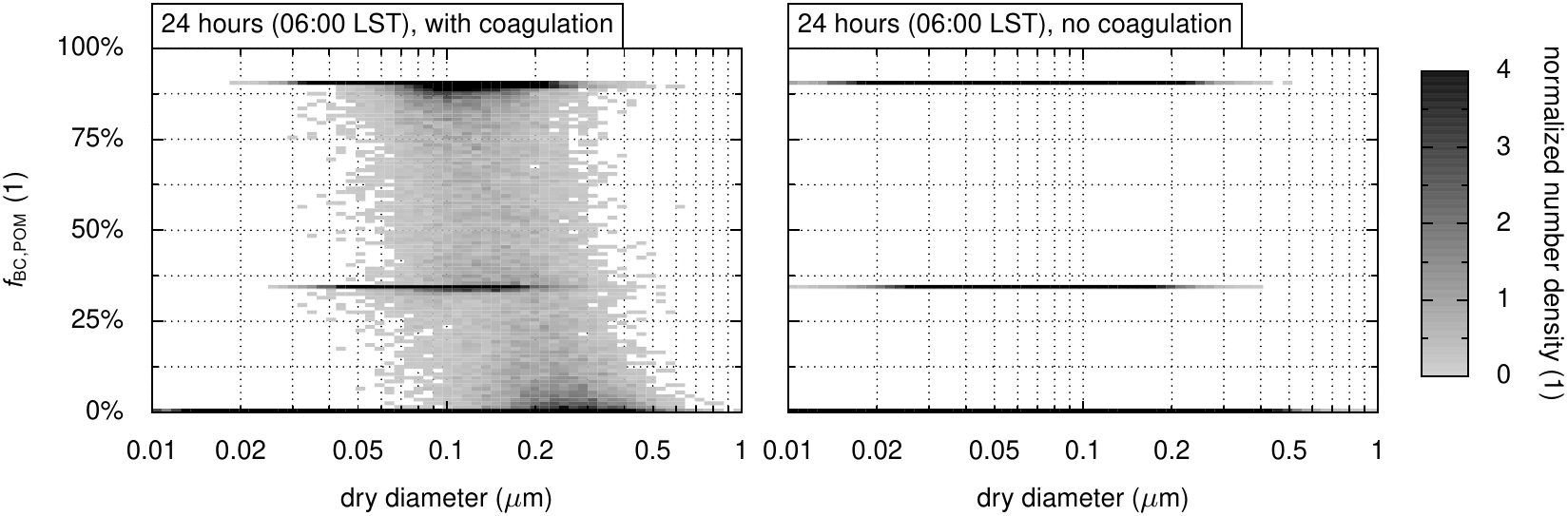}
  \end{center}
  \caption{\label{fig:aero_2d_oc} Normalized two-dimensional number
    distribution $\hat{n}_{{\rm BC},{\rm POM}}(f, D)$ after 24 hours of
    simulation (06:00 LST the following day), with and without
    coagulation. The maximum plotted value for $\hat{n}_{{\rm BC},{\rm
        dry}}(f, D)$ is capped at 4 to allow better resolution.}
\end{figure*}

The impact of coagulation on the mixing state with respect to the
primary components BC and POM is shown in
Figure~\ref{fig:aero_2d_oc}. The left figure displays the BC mixing
state with respect to POM, $f_{\rm BC,POM}$, after 24 hours with
coagulation. POM was emitted as a constituent of primary particles,
which can be seen as horizontal lines with high number densities at
$f_{\rm BC,POM}=90\%$ for diesel car emissions, $f_{\rm BC,POM}=34\%$
for gasoline car emissions and $f_{\rm BC,POM}=0\%$ for meat cooking
emissions. The mixing states between these could only form as a result
of coagulation. Since coagulation is most efficient between particles
of different size, we observe that these mixed particles
preferentially formed in a specific size range.  For sizes larger than
$D = 0.05 \rm \, \mu m$, POM/BC mixtures of various degree of mixing
formed due to coagulation.  Below $0.05 \rm \, \mu m$, coagulation
produced very few particles, so particles at these sizes were at their
initial BC/POM mixing state.  Nearly all particles below $0.03 \rm \,
\mu m$ were removed by coagulation.  For comparison the right panel of
Figure~\ref{fig:aero_2d_oc} shows the BC mixing state with respect to
POM, $f_{\rm BC,POM}$, at the end of the simulation without
coagulation.  For this case the intermediate mixing states did not
occur, and particles below $D = 0.03 \rm \, \mu m$ remained.

\begin{figure}
  \begin{center}
    \includegraphics{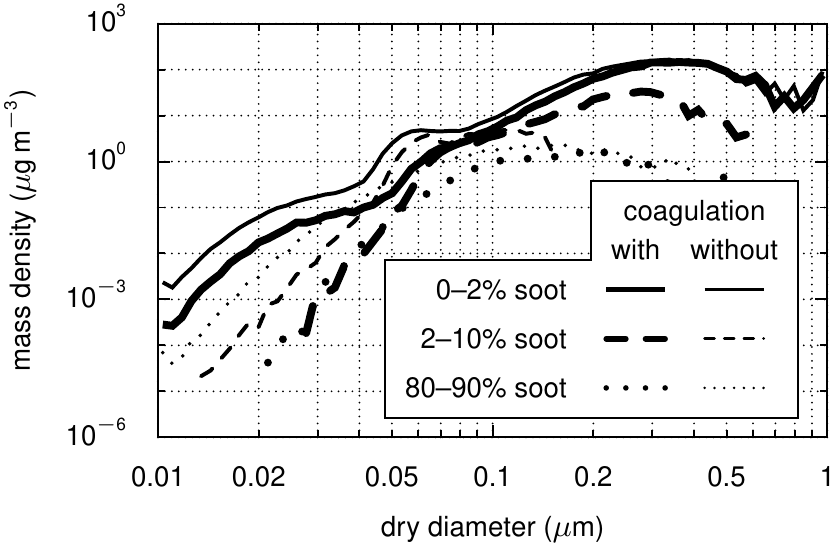}
  \end{center}
  \caption{\label{fig:aero_bc_mixing} Mass distribution after 24 hours
    (06:00~LST the following day) for three different mixing state
    ranges: $m_{{\rm BC},{\rm dry}}([0\%,2\%], D)$, $m_{{\rm BC},{\rm
        dry}}([2\%,10\%], D)$, and $m_{{\rm BC},{\rm
        dry}}([80\%,90\%], D)$, as defined in
    equation~(\ref{eqn:mass_den_range}). The cases with and without
    coagulation are plotted for comparison.}
\end{figure}

Figure~\ref{fig:aero_bc_mixing} shows the one-dimensional size
distributions of total mass density (including water) for different
ranges of mixing states at the end of the simulation, comparing the
cases with and without coagulation. From this we see that coagulation
did not simply reduce the number densities, but also shifted black
carbon mass within the diameter-$f_{\rm BC,dry}$ space. The mass of
particles smaller than $D = 0.05 \rm \, \mu m$ with high BC content
($f_{\rm BC,dry}$ between 80\% and 90\%) was reduced by 90\% due to
coagulation. The mass of particles smaller than $D = 0.05 \rm \, \mu
m$ with very low BC content ($f_{\rm BC,dry}$ between 0\% and 2\% BC)
was reduced by 81\% when coagulation was included. A very large
difference between the cases with and without coagulation occurred for
$f_{\rm BC,dry}$ between 2\% and 10\% and for the size range above $D
= 0.1 \rm \, \mu m$. Mass in this range of parameters arose mainly
from coagulation of large, BC-free particles with small BC-containing
particles and this mass increased by 1458\% when coagulation was
included.

\begin{figure}
  \begin{center}
    \includegraphics{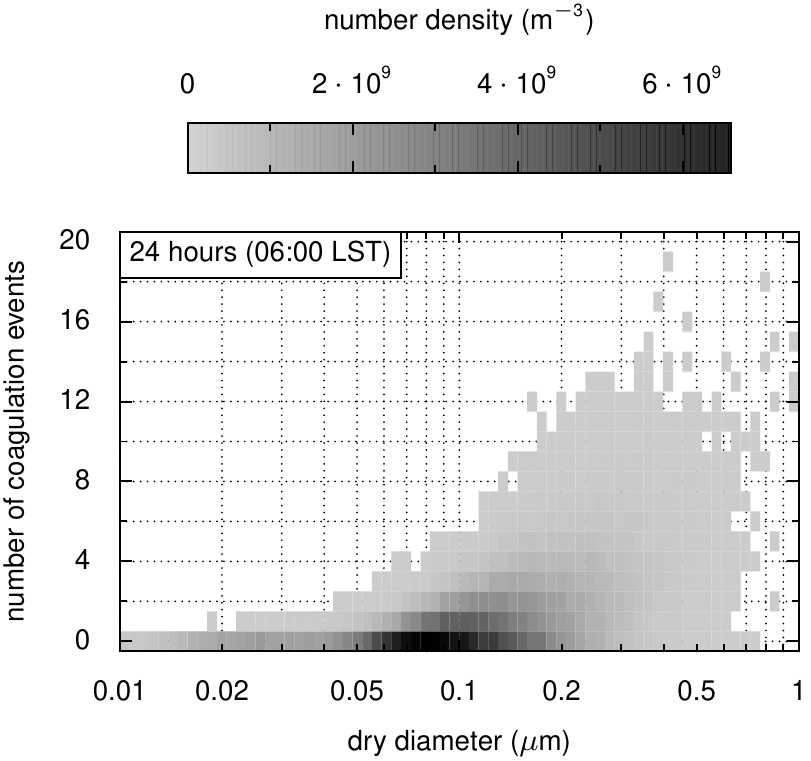}
  \end{center}
  \caption{\label{fig:aero_2d_n_orig} Two-dimensional number distribution
    $n_{\rm coag}(k,D)$ showing the number of coagulation events
    experienced after 24 hours (06:00 LST the following day), as
    defined in equation~(\ref{eqn:n_coag}). The maximum plotted value
    for $n_{\rm coag}(k, D)$ is capped at $6\cdot10^9\,\rm{m^{-3}}$ to
    allow better resolution.}
\end{figure}

With PartMC it is straightforward to track the number of coagulation
events experienced by the individual particles.
Figure~\ref{fig:aero_2d_n_orig} shows the two-dimensional number
distribution $n(k,D)$, with $k$ being the number of coagulation
events. At the end of the simulation, 5\% of particles had undergone
at least 5 coagulation events. The largest number of coagulation
events was 19 and occurred for particles on the right hand side of the
spectrum with sizes larger than $D = 0.4 \rm \, \mu m$. Given a
certain size, the number of coagulation events varied, which shows the
stochastic nature of the coagulation process. The range of variation
was greater for larger particles. For example, while the number of
coagulation events varied between 0 and 5 for a $D = 0.1 \rm \, \mu m$
particle, it ranged between 0 and 12 for a $D = 0.3 \rm \, \mu m$
particle.

\section{Summary}

In this paper we presented the development and the application of a
stochastic particle-resolved aerosol model, PartMC-MOSAIC. It
explicitly resolves the composition of individual particles in a given
population of different types of aerosol particles, and accurately
tracks their evolution due to emission, dilution, condensation and
coagulation. To make the direct stochastic particle-based method
practical, we implemented an accelerated coagulation method. With this
method we achieved optimal efficiency for applications when the
coagulation kernel is highly non-uniform, as is the case for many
realistic environments. The highly accurate treatment of aerosol
dynamics and chemistry makes PartMC-MOSAIC suitable for use as a
numerical benchmark of mixing state for more approximate models. The
current version of PartMC is available under the GNU General Public
License (GPL) at http://www.mechse.uiuc.edu/research/mwest/partmc/,
and the MOSAIC code is available upon request from R.~A.~Zaveri.

PartMC-MOSAIC was applied to an idealized example urban plume case to
simulate the evolution of urban aerosols of different types due to
coagulation and condensation, focusing on the aging process of BC. For
the first time results of the aerosol composition and size
distribution are available as a fully multi-dimensional size
distribution without any a priori assumptions about the mixing
state. This detail of information was only achievable with a
particle-resolved model.

To display the results, we projected the multi-dimensional size
distributions to two-dimensional distributions depending on particle
size and BC mass ratio. We specifically discussed the results for BC
mass ratios defined with respect to all other dry constituents,
$f_{{\rm BC},{\rm dry}}$, and to POM, $f_{{\rm BC},{\rm POM}}$. Due to
the diurnal variations in temperature, relative humidity, and gas
phase concentrations, the thermodynamic equilibrium conditions for the
ammonium-sulfate-nitrate system changed continuously. The aerosol
hydration hysteresis effect led to the co-existence of metastable
(wet) and stable (dry) particles in the air parcel during the daytime,
depending on their time of emission. Since the formation of ammonium
nitrate depends on the particle phase state, this in turn resulted in
pronounced differences in how the aging proceeded. As a result of
coagulation and condensation, after 24~hours of simulation, the
aerosol population evolved into a state where a continuum of BC mixing
states existed. Coagulation was effective in removing smaller
particles, reducing number densities by 84\%, 66\%, and 29\% at
diameters of $0.05$, $0.07$, and $0.10\rm\,\mu m$ in the example
simulation.

\begin{acknowledgments}
Funding for N.~Riemer and M.~West was provided by the National Science
Foundation (NSF) under grant ATM 0739404.

Funding for R.~A.~Zaveri and R.~C.~Easter was provided by the
Aerosol-Climate Initiative as part of the Pacific Northwest National
Laboratory (PNNL) Laboratory Directed Research and Development (LDRD)
program. Pacific Northwest National Laboratory is operated for the
U.S. Department of Energy by Battelle Memorial Institute under
contract DE-AC06-76RLO 1830.
\end{acknowledgments}

\bibliographystyle{agu04}
\bibliography{refs}

\end{article}


\end{document}